\documentclass[aps,twocolumn,groupedaddress,amsfonts]{revtex4}
\usepackage{amsmath,amssymb,natbib,bm,graphicx,psfrag,ulem}
\usepackage{times}
\newcommand{\Rey}{\mbox{\rm Re}}            
\newcommand{\Rm}{R_\mathrm{m}}              
\newcommand{\Rmt}{\tilde{R}_\mathrm{m}}              

\newcommand{\pderiv}[2]{\frac{\partial #1}{\partial #2}} 

\begin{document}

\title{Fluctuation dynamo based on magnetic reconnections}
\author{Andrew~W.~Baggaley}
\email{a.w.baggaley@ncl.ac.uk}
\author{Carlo~F.~Barenghi}
\author{Anvar~Shukurov}
\affiliation{School of Mathematics and Statistics, Newcastle University,
Newcastle upon Tyne, NE1 7RU, UK}
\author{Kandaswamy Subramanian}
\affiliation{Inter-University Centre for Astronomy and Astrophysics, Post Bag 4,
Ganeshkhind, Pune 411 007, India}


\begin{abstract}
We develop a new model of the fluctuation dynamo in which the magnetic field
is confined to thin flux ropes advected by a multi-scale flow which models
turbulence. Magnetic dissipation occurs only via reconnections of flux ropes.
The model is particularly suitable for rarefied plasma, such as the Solar
corona or galactic halos. We investigate the kinetic energy release into heat,
mediated by dynamo action, both in our model and by solving the induction
equation with the same flow. We find that the flux rope dynamo is more than an order of
magnitude more efficient at converting mechanical energy into heat. The
probability density of the magnetic energy released during reconnections has a
power-law form with the slope $-3$, consistent with the Solar corona heating
by nanoflares.
We also present a nonlinear extension of the model. This shows that a plausible
saturation mechanism of the fluctuation dynamo is the suppression of turbulent
magnetic diffusivity, due to suppression of random stretching
at the location of the flux ropes.
We confirm that the probability distribution function of the magnetic
line curvature has a power-law form suggested by \citet{Sheck:2002b}. We
argue, however, using our results that this does not imply a persistent folded
structure of magnetic field, at least in the nonlinear stage.
\end{abstract}

\maketitle

\section{Introduction}
Dynamo action, i.e., the amplification of magnetic field by
the motion of an electrically conducting fluid (plasma), is the most likely
explanation for the omnipresence of astrophysical magnetic fields.
Ohmic dissipation, however small, is essential in
order to achieve the development of the dynamo eigensolutions and to smooth
out the spatial variations of the magnetic field.
The evolution of the magnetic field $\mathbf{B}$
embedded in a velocity field $\mathbf{u}$
is governed by the following closed equation:
\begin{equation}\label{induction}
  \pderiv{\mathbf{B}}{t}=\nabla \times (\mathbf{u} \times \mathbf{B})
        +\widehat{\cal L}\mathbf{B},
\end{equation}
where $\widehat{\cal L}$ is an operator describing the magnetic dissipation.

In rarefied astrophysical plasmas, such as the Solar co\-ro\-na, hot gas in spiral and
elliptical galaxies, galactic and accretion disc halos, and laboratory
plasmas, an important (or even dominant) mechanism for the dissipation of
magnetic field is the reconnection of magnetic lines rather than magnetic
diffusion \citep{priest:2000}, the latter modelled with $\widehat{\cal
L}=\eta\nabla^2$ (if $\eta=\mbox{const}$). Discussions of astrophysical
dynamos often refer to magnetic reconnection, but attempts to include
features specific of magnetic reconnection into dynamo models are very rare
\citep[see however ][]{Blackman:1996}. On the other hand, theories of magnetic
reconnection (and the resulting estimates of the plasma heating rate) rarely,
if ever, refer to the dynamo action as the widespread mechanism maintaining
magnetic fields. This paper attempts to bridge the gap between the two major
topics of astrophysical magnetohydrodynamics (dynamos and reconnections) by
developing a dynamo model which explicitly incorporates magnetic
reconnections.

\begin{figure}[h!]
  \begin{center}
    \psfrag{x}[c]{$k$}
    \psfrag{y}[c][l]{$\widehat{\cal L}_k$}
    \psfrag{t}[l]{${2\pi}/d_0$}
    \psfrag{u}[c]{$k^2$}
    \psfrag{v}[c]{$k^4$}
    \includegraphics[width=0.4\textwidth]{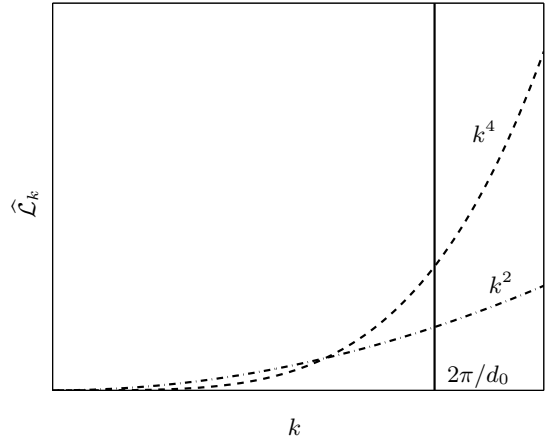}
    \caption{\label{fourier}  A schematic representation of the magnetic
dissipation operator $\widehat{\cal L}$ in Fourier space: usual diffusion
$\widehat{\cal L}_k \propto k^2$ (dash-dotted), hyperdiffusion $\widehat{\cal
L}_k \propto k^4$ (dashes) and reconnections at a scale $d_0$ as described by
our model (solid).}
\end{center}
\end{figure}

The nature of the dissipation mechanism is important for the dynamo action as
it affects the growth time of magnetic field, its spatial form and the rate of
plasma heating by the electric currents. For example, dynamo action with
hyperdiffusion, $\widehat{\cal L}=-\eta_1\nabla^4$ (and with a helical
$\mathbf{u}$) has larger growth rate and stronger steady-state magnetic fields
than a similar dynamo based on normal diffusion \citep{Brandenburg:2002}. This
is not surprising, as the hyperdiffusion operator, having the Fourier
dependence of $k^4$, rather than the $k^2$ dependence of normal diffusion, has
weaker magnetic dissipation at larger scales as shown in Fig.~\ref{fourier}.
This allows the magnetic field to grow unimpeded by dissipation as magnetic
dissipation is confined to relatively small regions. The release of magnetic
energy in smaller regions (and larger current densities) in hyperdiffusive
dynamos may also lead to a higher rate of conversion of kinetic energy to heat
via magnetic energy. One of the aims of this paper is to demonstrate that this
statement is especially true in the case of magnetic reconnections.

Magnetic hyperdiffusion also appears in the context of continuous models of
self-organised criticality in application to the heating of the Solar corona
\citep{SOC}. The aim of such models is to reproduce the observed frequency
distribution of various flare energy diagnostics. As we show here, our model
exhibits a power-law probability distribution of the magnetic energy release
similar to that observed in the Solar corona.

Magnetic reconnection may correspond to an even more extreme form of the
dissipation operator than the hyperdiffusion: here magnetic flux tubes
dissipate their energy only when in close contact with each other, so that the
Fourier transform of $\widehat{\cal L}$ should be negligible at all scales
exceeding a certain reconnection length $d_0$ (see Fig.~\ref{fourier}). It is
then natural to expect that dynamos based on reconnections (as opposed to
those involving magnetic diffusion) will exhibit faster growth of magnetic
field, more intermittent spatial distribution and stronger plasma heating. In
this paper we consider dynamo action based on direct modelling of magnetic
reconnections. For this purpose, we follow the evolution of individual closed
magnetic loops in various flows (known to support dynamo action) and reconnect
them directly whenever they come into a sufficiently close contact,
with appropriate magnetic field directions. First results of our simulations
can be found in \citep{BBSS09}.

\begin{figure}
  \begin{center}
    \includegraphics[width=0.35\textwidth]{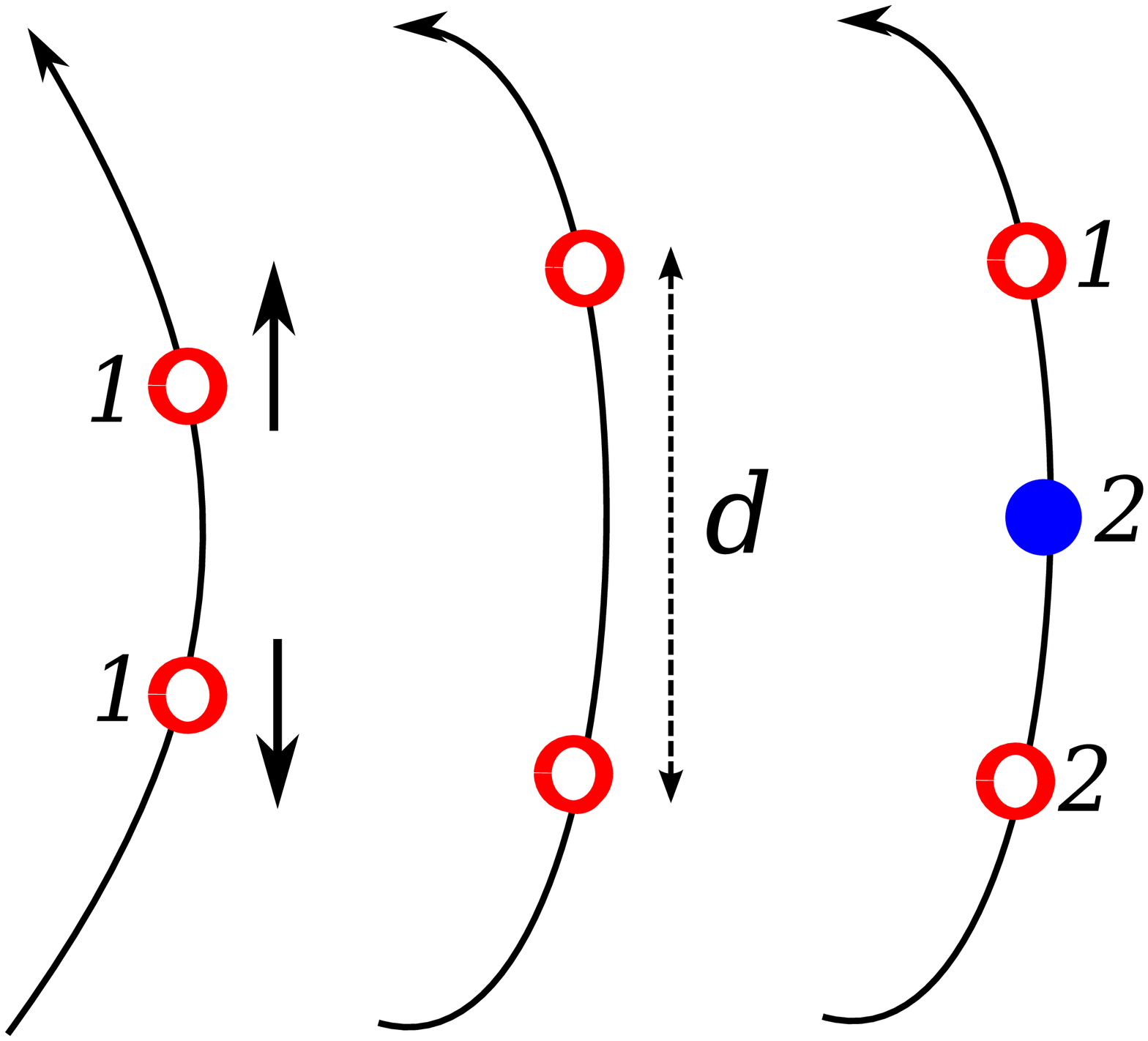}
    \caption{\label{insert_scheme} (Colour online) The algorithm for inserting new trace
particles in a stretched (left to right) or contracting (right to left) magnetic flux tube.
 If the distance between any two
trace particles (shown with red/open circles) exceeds a length scale
$d$, a new particle is inserted between the two particles shown with a
blue/filled circle. The label at each particle represents magnetic field strength at
that location
}
  \end{center}
\end{figure}

\section{The flux rope model}
We model the magnetic field by considering the evolution of thin flux tubes,
frozen into a flow, each with constant magnetic flux $\psi$. We first focus on
the kinematic behaviour, where the velocity field is independent of magnetic
field. Later we shall introduce the Lorentz force into the system to account
for the back reaction of the magnetic field on the velocity field. To ensure
that $\nabla \cdot \mathbf{B}=0$, we require that our flux tubes always take
the form of closed loops. Numerically, we disctretize the loops into fluid
particles and track their positions and relative order (i.e., magnetic field
direction) by introducing a flag $P$, whose value increases along a given
magnetic flux tube. Initially the particles are set a small distance apart,
$0.75d$, where $d$ is a certain (small) constant length scale. If, during the
evolution of the loops, the distance between two neighbouring fluid particles
on a loop becomes larger than $d$, we introduce a new particle between them,
as illustrated in Fig.~\ref{insert_scheme}. We use linear interpolation to
place the new particle halfway between the old ones. For example if inserting
a new particle $\mathbf{x}_\mathrm{c}$ between particles $\mathbf{x}_\mathrm{a}$
and $\mathbf{x}_\mathrm{b}$ the position of the new particle is given by,
\begin{equation}\label{lin_interp}
\mathbf{x}_\mathrm{c}=
{\textstyle\frac{1}{2}}(\mathbf{x}_\mathrm{b}+\mathbf{x}_\mathrm{a}).
\end{equation}
The separation between the new particles is thus greater than $0.5d$ -- a
feature which will be important when we consider removing particles. The
effective spatial resolution of our model is thus close to $d$. We shall
discuss later a prescription for $\mathbf{x}_\mathrm{c}$ which is more
accurate than Eq.~(\ref{lin_interp}).

Each particle is also assigned a flag $B$ which denotes the strength of the
magnetic field at that position on the loop. Assuming magnetic flux conservation
and incompressibility, the magnetic field strength in the flux tube is
proportional to its length. Initially the magnetic field is constant at all
particles in each loop, $B=1$. However, when a new particle is introduced, the
magnetic field is doubled at certain particles, as shown in
Fig.~\ref{insert_scheme}. Importantly, the field strength is increased at two
out of three particles involved: this prescription emerged from our
experimentation with various schemes, and allows us to reproduce the evolution
of magnetic field strength in a shear flow. Conversely, when the flow reduces
the separation of particles to less than $0.5d$, we remove a particle. The
value of the magnetic field strength flag is also halved on the remaining
particles in a manner consistent with the above algorithm. We have verified
that this prescription reproduces accurately an exact solution of the
induction equation for a simple shear flow.

Results presented below have been obtained with a typical number of trace
particles of order $10^4$.

\begin{figure}
  \begin{center}
    \psfrag{t}{$t$}
    \psfrag{$|B|$}[c]{$|\mathbf{B}|$}
    \includegraphics[width=0.4\textwidth]{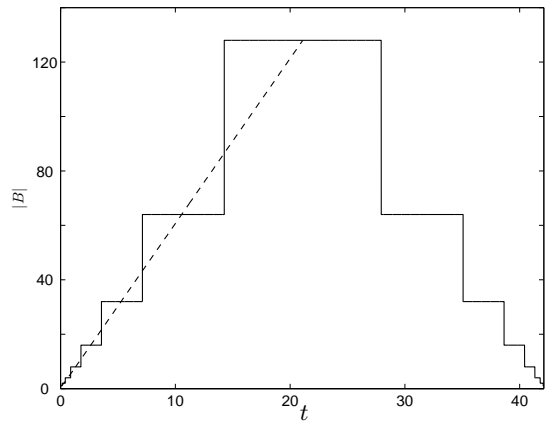}
    \caption{\label{line2} $|\mathbf{B}|$ at a specific position
    ($y=1.$) in a shear flow (\ref{gaussian_shear}) whose velocity is reversed at $t\approx20$. The dotted
line shows the analytic solution (\ref{gauss_ind}), and numerical results are
shown with solid line. The initial field strength is $B=1$.}
\end{center}
\end{figure}

\begin{figure}
  \begin{center}
    \psfrag{x}{$x$}
    \psfrag{y}{$y$}
    \includegraphics[width=0.4\textwidth]{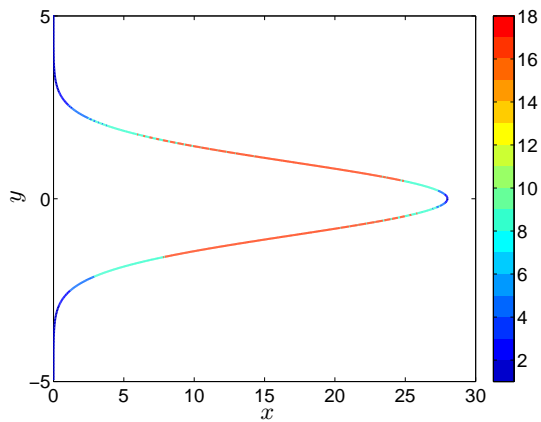}
    \caption{\label{shear_colour}(Colour online) The shape of the flux tube
stretched by the flow, given in Eq.~(\ref{gaussian_shear}) at $t=2.7$. Colour
coding shows the magnetic field strength according the key (right), $B_0=1$.}
  \end{center}
\end{figure}

\begin{figure}
  \begin{center}
    \includegraphics[width=0.4\textwidth]{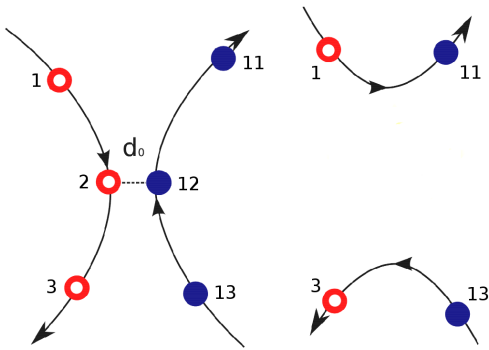}
    \caption{\label{recon_scheme}(Colour online) Reconnection occurs when the
 distance between two trace particles reduces to $d_0$ (left); the connection
of the particles on  a magnetic flux tube changes after the reconnection
(right).}
\end{center}
\end{figure}

\begin{figure}
  \begin{center}
    \includegraphics[width=0.25\textwidth]{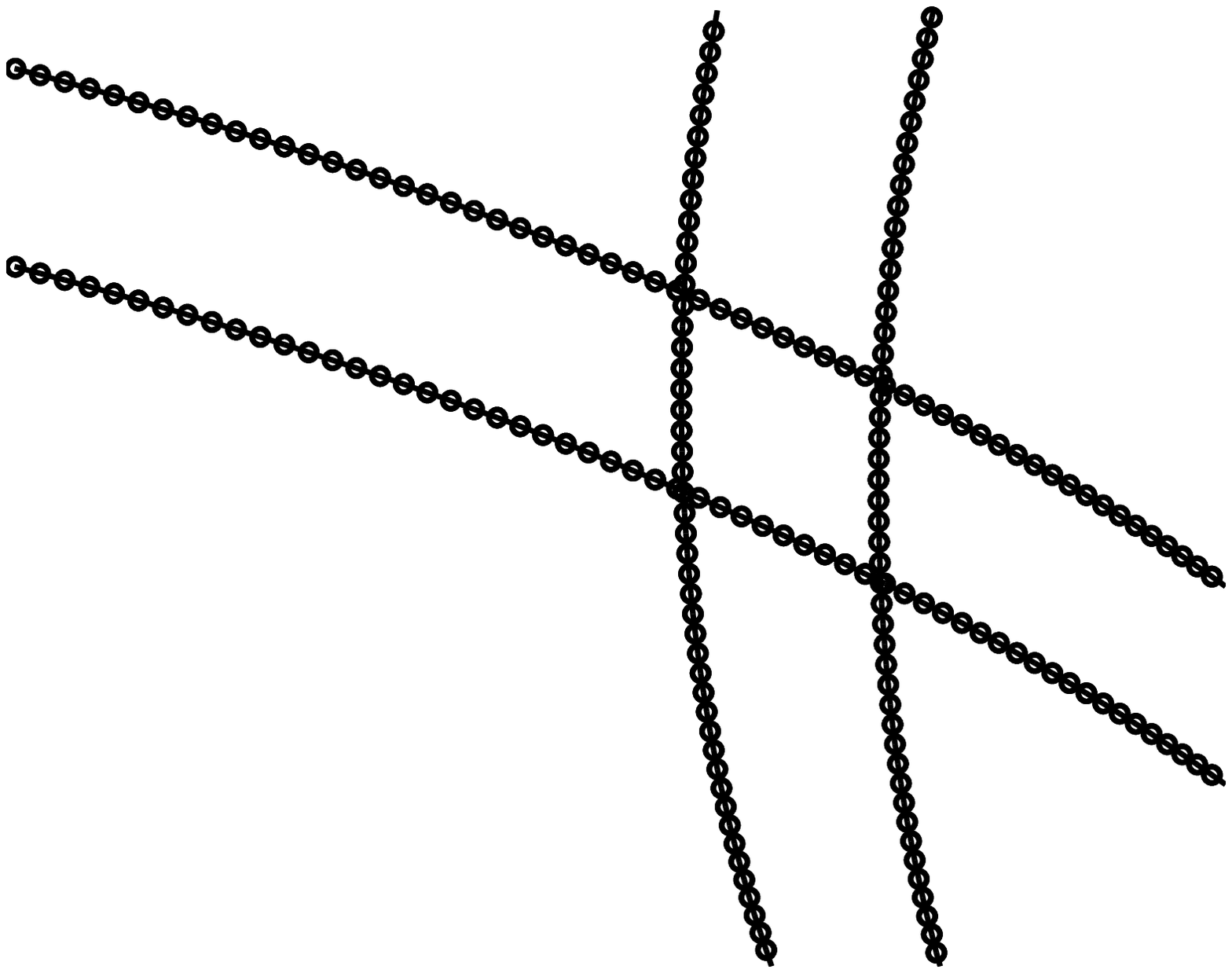}
    \includegraphics[width=0.25\textwidth]{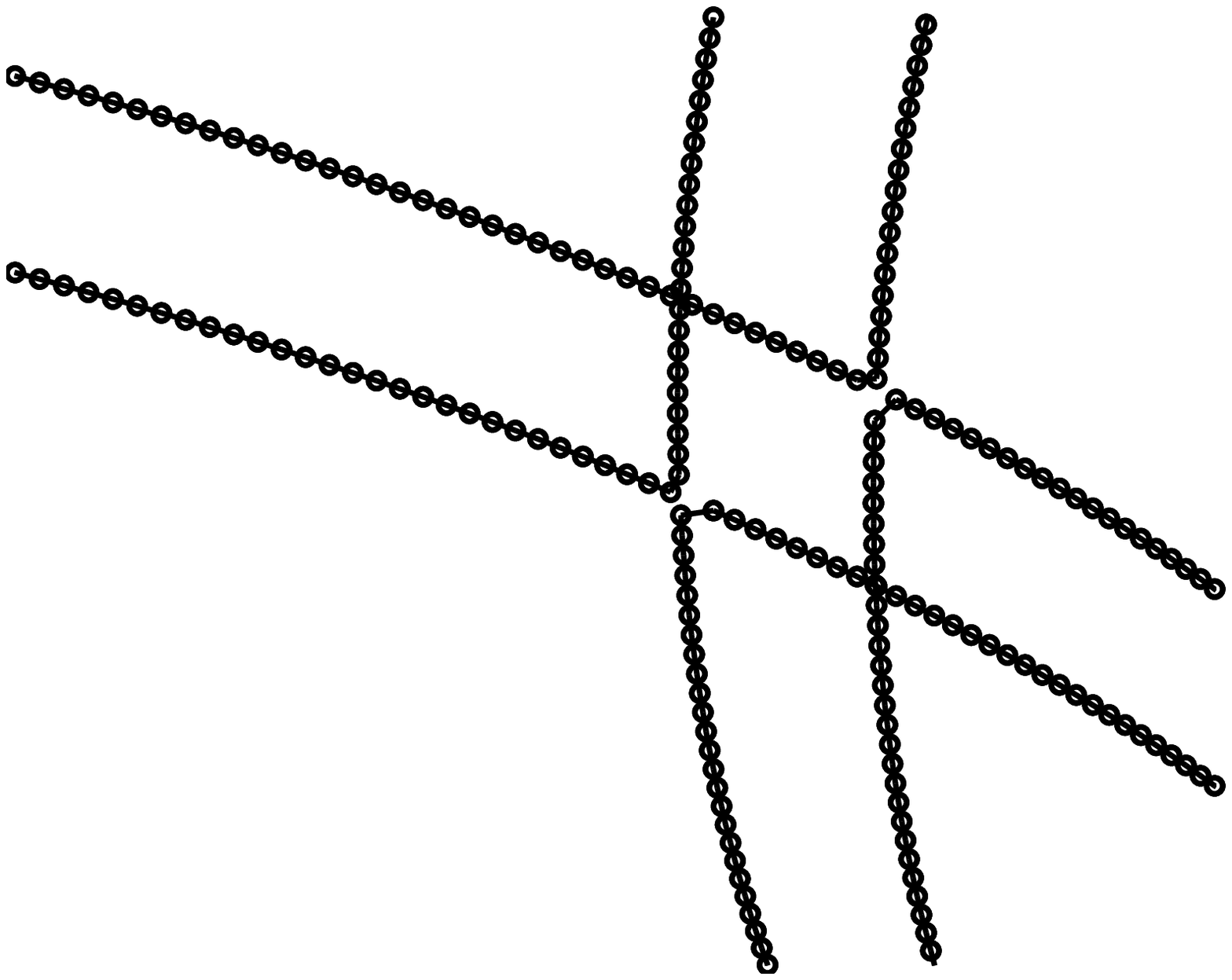}
    \caption{\label{recon_snapshot}Snapshots of two simultaneous reconnection
events, before the reconnection (top) and after (bottom). Note the change of
connections of the flux ropes after the reconnection.}
\end{center}
\end{figure}

\begin{figure}
  \begin{center}
    \psfrag{x}{$x$}
    \psfrag{y}{$y$}
    \psfrag{t1}{t=5.0}
    \psfrag{t2}{t=40.0}
    \psfrag{t4}{t=46.5}
    \psfrag{t3}{t=200.0}
    \includegraphics[width=0.26\textwidth,height=0.33\textwidth]{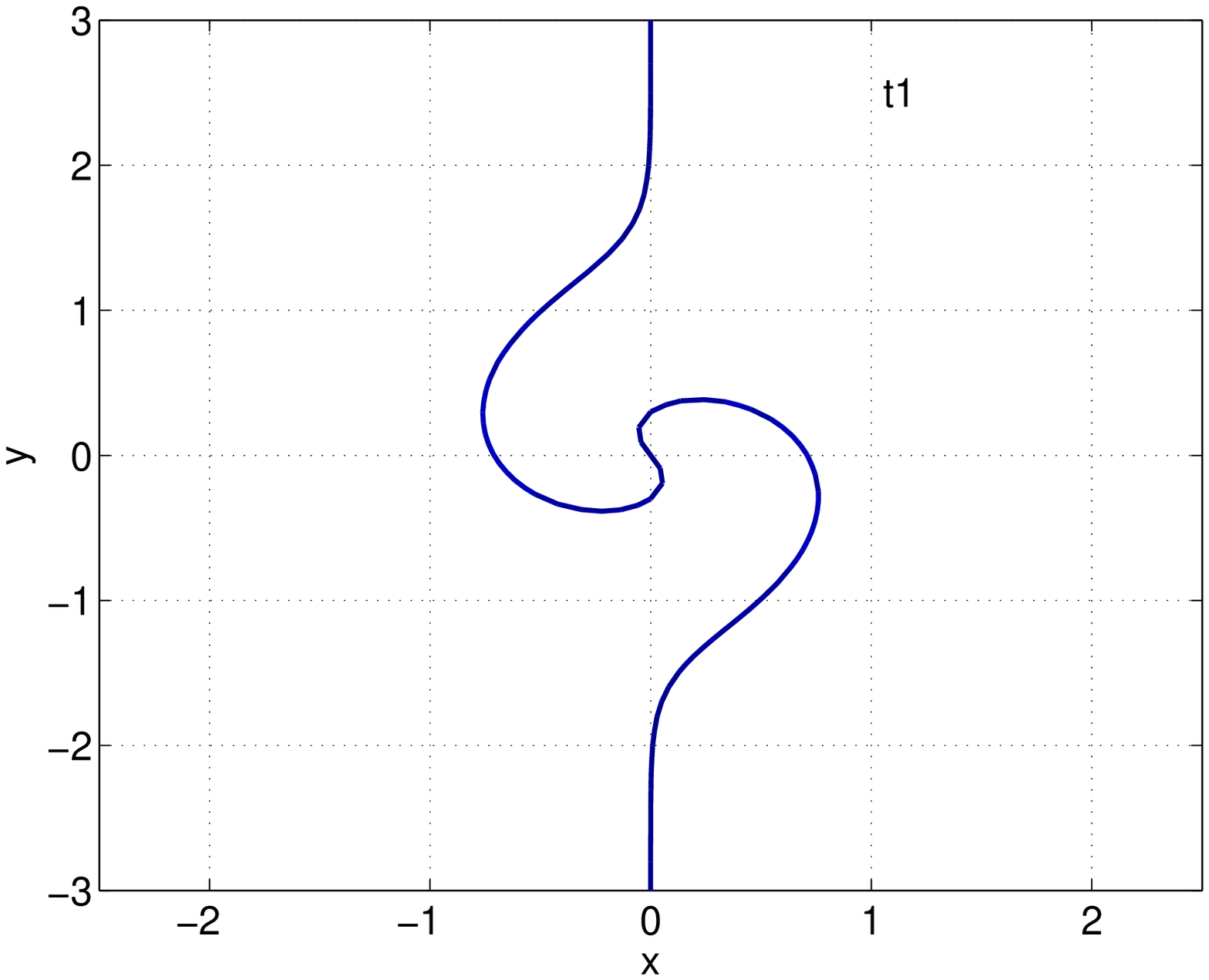}\\
\mbox{}\quad\ \ \ \includegraphics[width=0.24\textwidth,height=0.29\textwidth]{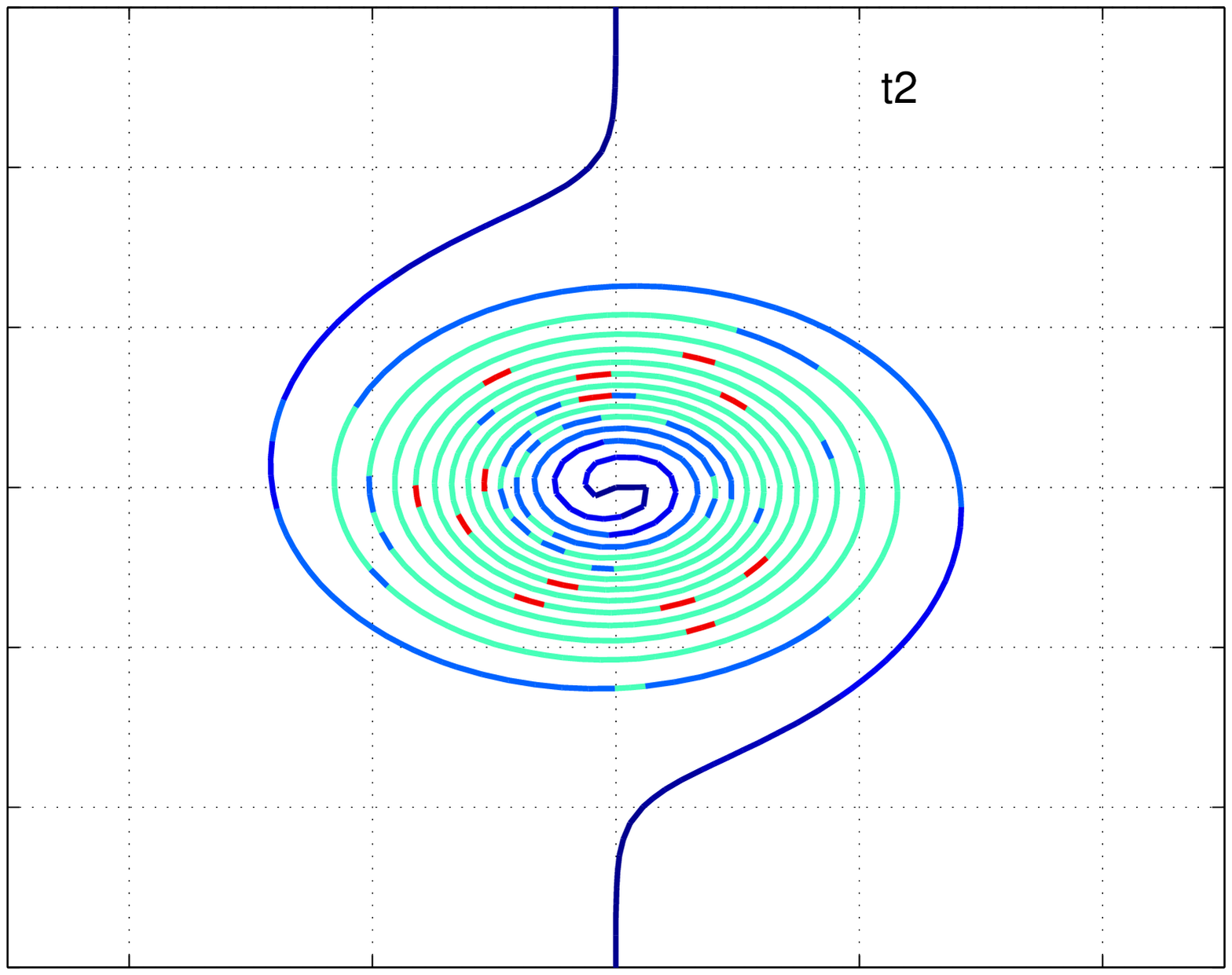}\\
\mbox{}\quad\ \ \ \includegraphics[width=0.24\textwidth,height=0.29\textwidth]{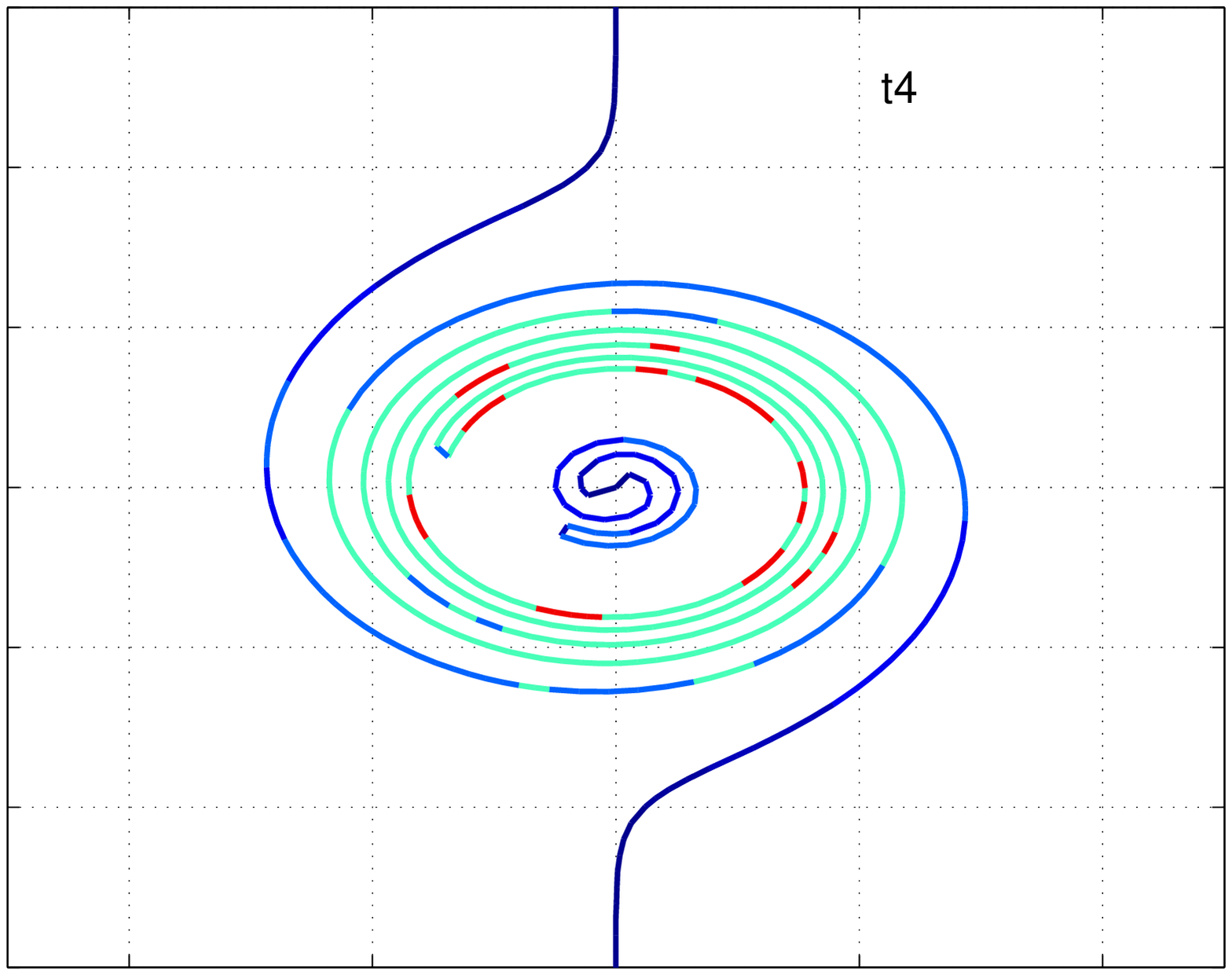}\\
\mbox{}\quad\quad\quad\ \ \ \includegraphics[width=0.27\textwidth,height=0.29\textwidth]{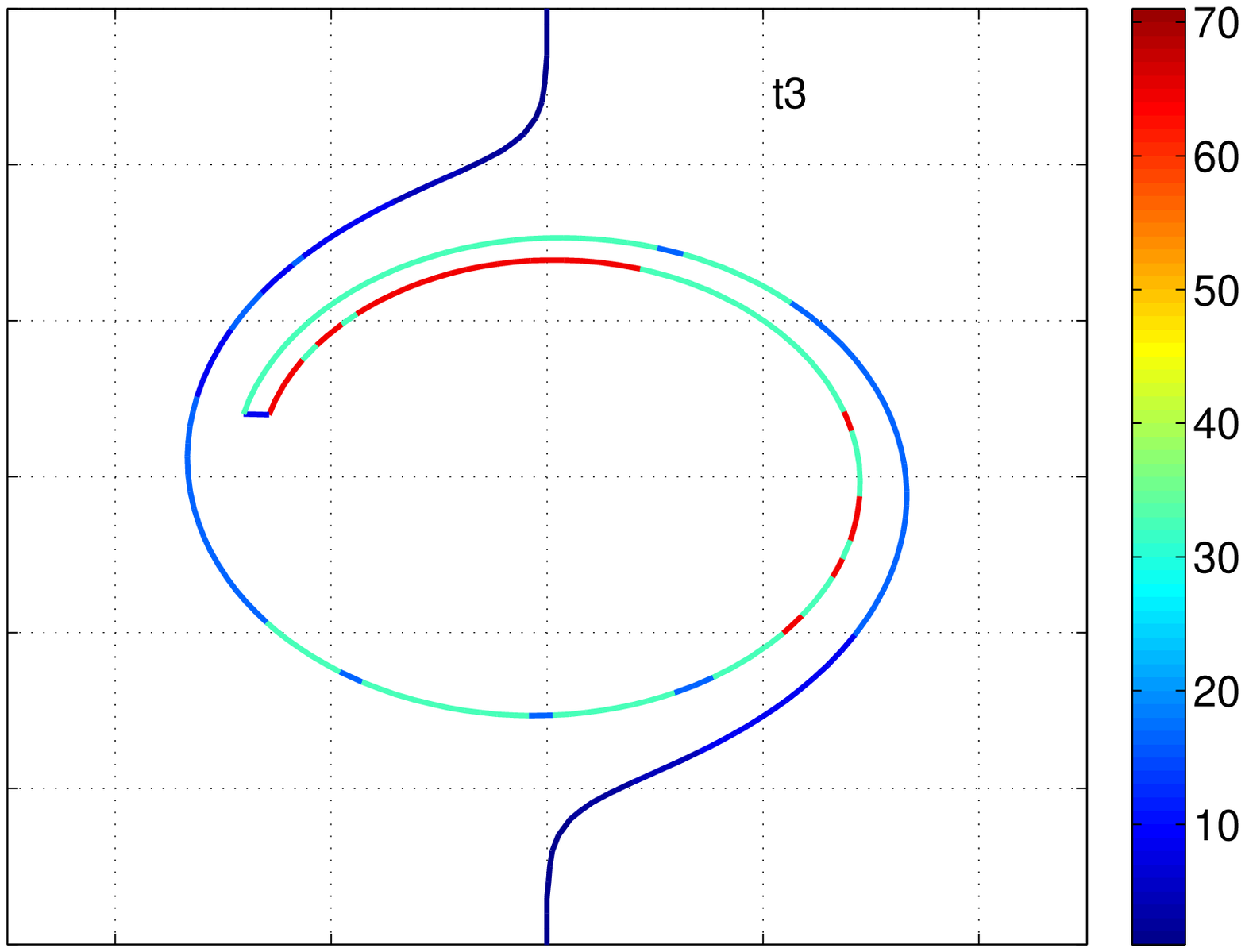}\\
    \caption{\label{flux_expuls_snaps}(Colour online) Flux expulsion by differential rotation
(\ref{drot}): the form of magnetic line initially aligned with the
$y$-axis at different times indicated in the corner of each frame.
The field strength grows as the magnetic line is wound around by the
differential rotation. Eventually the separation of neighbouring turns becomes
less than $d_0$ and reconnections destroy
the field. Magnetic field strength is colour coded as
in Fig.~\protect{\ref{shear_colour}}.}
\end{center}
\end{figure}

\begin{figure}
  \begin{center}
    \psfrag{x}[c][position=1mm]{\vspace{3mm} $t$}
    \psfrag{y}[c][position=1mm]{\vspace{1mm} ${B}_\mathrm{rms}$}
    \includegraphics[width=0.4\textwidth]{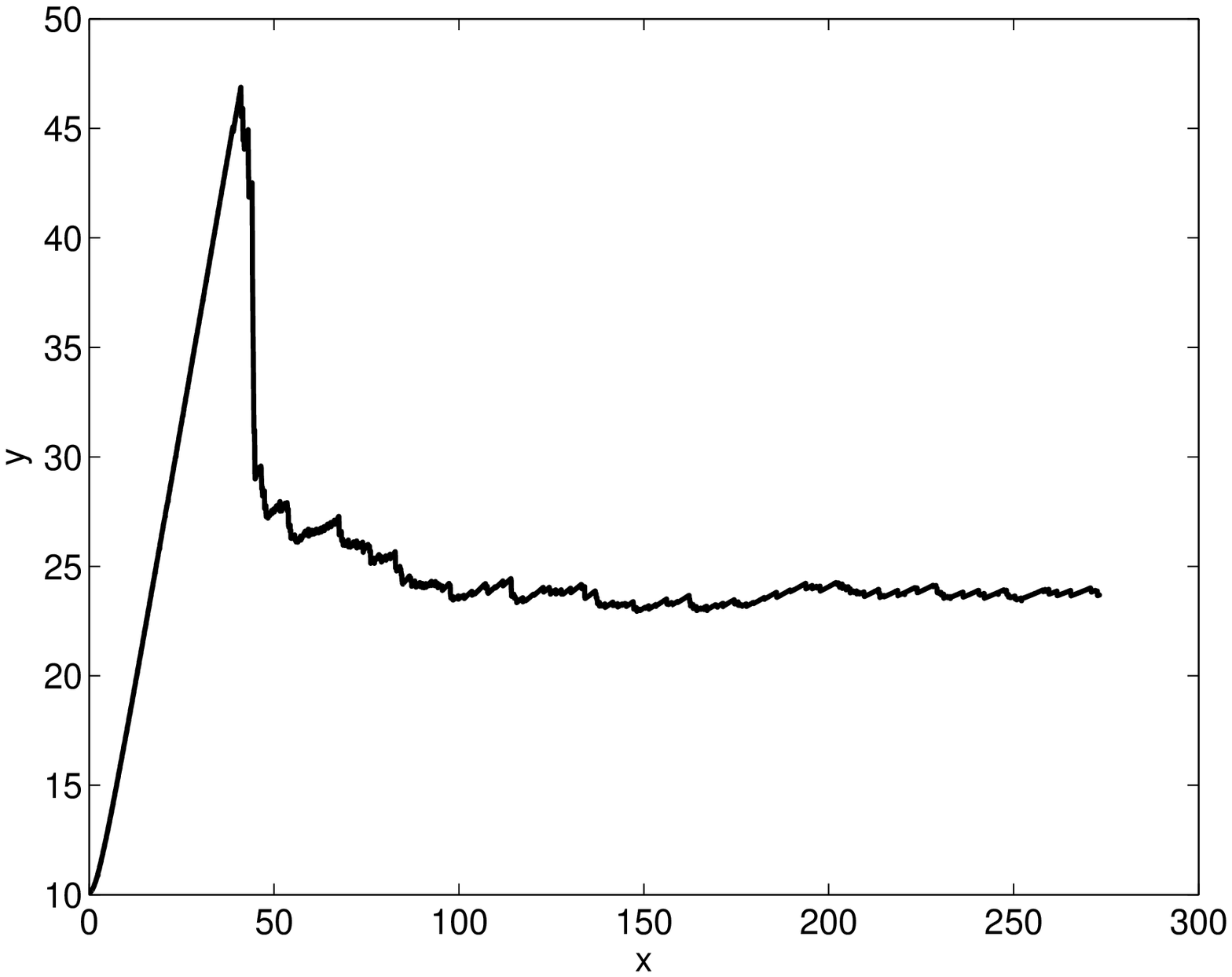}
    \psfrag{x}[c][position=1mm]{\vspace{1mm} $\log \Rmt$}
    \psfrag{y}[c][position=1mm]{\vspace{1mm} $\log {B}_\textrm{rms, max}$}
    \includegraphics[width=0.4\textwidth]{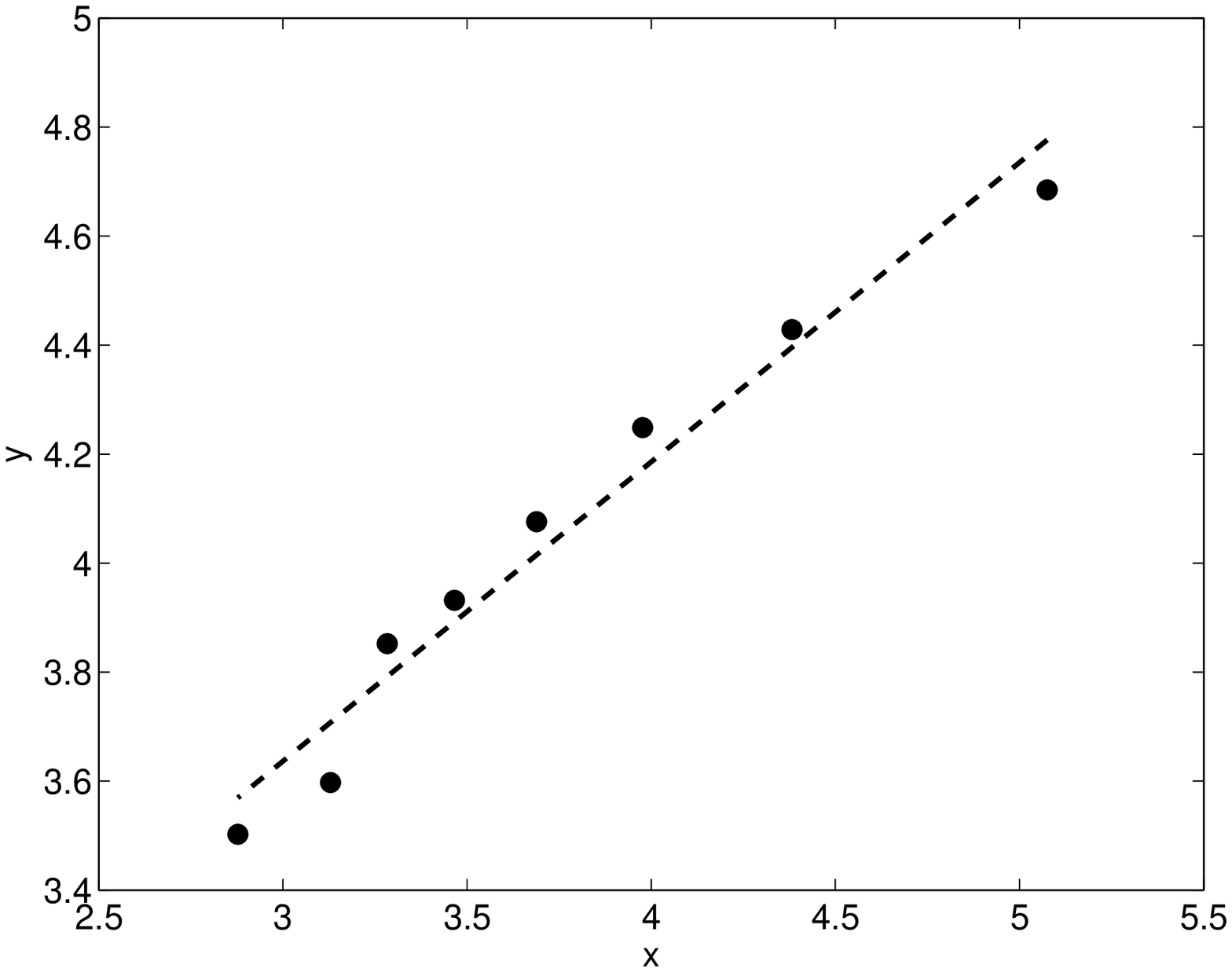}
    \caption{\label{flux_expuls_analysis} The upper panel shows
    the root-mean-square magnetic field strength ${B}_\textrm{rms}$ as a function
of time for the simulation shown in
    Fig.~\protect\ref{flux_expuls_snaps}. The lower panel represents
the scaling of the maximum values of ${B}_\textrm{rms}$ among eight simulations
    with decreasing $d_0$. The line of best fit, shown dotted, has the slope
    $0.54 \pm 0.21$}
  \end{center}
\end{figure}

\subsection{Shear flow test}
In order to test our model we consider a two dimensional shear flow
with a Gaussian profile,
\begin{equation}\label{gaussian_shear}
\mathbf{u}=(u_x,0)\;,
\quad
u_x=u_0 e^{-y^2/2},
\end{equation}
and a flux tube extended across the flow from $y=-\infty$ to $+\infty$. For
$\widehat{\cal L}\mathbf{B}=0$, Eq.~(\ref{induction}) can easily be solved
exactly to yield,
\begin{equation}\label{gauss_ind}
|\mathbf{B}|=B_0 \sqrt{1+u_0y^2e^{-y^2}t^2},
\end{equation}
where $B_0$ is the initial field strength.
Since,
\begin{equation}
\int_V |\mathbf{B}|\,dV=\int_{-\infty}^{\infty}\psi(l)\,dl \propto {L},
\end{equation}
where $\psi=\mbox{const}$ is the magnetic flux and $L$ is the length of the flux tube,
and since $\mathbf{B}$ is independent of $x$, we have
\begin{equation}\label{Lpropto}
{L} \propto B_0 \int_{-\infty}^{\infty} \sqrt{1+u_0y^2e^{-y^2}t^2} \, dy.
\end{equation}
We find excellent agreement between Eq.~(\ref{Lpropto}) and our numerical solution.
In Fig.~\ref{line2} we plot $|\mathbf{B}|$ at a fixed value of $y$ versus time
to compare it with Eq.~(\ref{gauss_ind}). The comparison is quite
satisfactory; the step-wise change in the numerical solution for $B$ arises
because, in this simple flow with a the shear rate slowly varying in space,
many new trace particles are introduced simultaneously as the flux tube is
stretched, and then no particles are added for some time until the next series
of particle insertions. After $t \approx 20$, we reverse the flow to observe
that the particles and magnetic field return to their initial states, to
demonstrate that our algorithm correctly describes the contraction of the flux
tubes as well. Fig.~\ref{shear_colour} shows that the flux tube adopts the
shape of the flow before the flow field is reversed, colour coding indicating
the magnetic field strength.

\subsection{Reconnections}
Reconnections are introduced into the model in a straightforward manner. If
the separation between two particles, which are not neighbours, becomes less
than a certain scale $d_0$, we reconnect their associated flux tubes by
reassigning the flags $P$ which identify the particles ahead and behind those
involved in the reconnection, as shown in Fig.~\ref{recon_scheme}. We found
that $d_0$ has to be comparable to the separation of the trace particles, $d$,
in order to obtain meaningful numerical results, e.g., $d_0=1.5d$. Two
particles are removed from the system after each reconnection event, those
labelled $P=2$ and $P=12$ in Fig.~\ref{recon_scheme}, and their magnetic
energy is lost, presumably to heat. We also monitor the cross product of the
tube tangent vectors close to the reconnection point. By ensuring that the
magnitude of the cross product is smaller than some tolerance $\epsilon\approx
10^{-2}$ and that the magnetic fields in the reconnecting loops are (almost)
oppositely directed, we prevent parallel flux tubes from reconnecting. In
Fig.~\ref{recon_snapshot} we show snapshots from a simulation before and after
two simultaneous reconnection events. Since we monitor the amount of magnetic
energy released in each reconnection event, we know the total magnetic energy
released by the reconnections over any given time period. Finally we introduce
a minimum loop size of $3d$, i.e., no magnetic loop can contain less than
three particles. Any smaller loop is removed from the system, releasing its
energy. We shall see later that this cutoff is important when we take
derivatives along the loops to calculate magnetic tension.

\subsection{Flux expulsion}\label{flux_expuls_section}
We test the reconnection algorithm by considering magnetic flux expulsion from
a region with closed streamlines \citep{Moffatt:1978}. Consider an initially
uniform magnetic field $\mathbf{B}_0$, in our case a single flux tube extended
over $-\infty<y<\infty$ along $x=0$. Differential rotation is applied to the
field, with velocity given in cylindrical polar coordinates by
\begin{equation}\label{drot}
\mathbf{u}=(u_r,u_{\theta}),
\quad
u_r=0,
\quad
u_{\theta}=\frac{1}{\sqrt{2\pi}}\exp\left(-\frac{r^2}{2\sigma^2}\right).
\end{equation}
Solutions of the induction equation grow linearly in time until a maximum
magnetic field is achieved,
\begin{equation}\label{Moffatt_flux}
|\mathbf{B}|_{\textrm{max}}=O(\Rm^{1/2})\mathbf{B}_0,
\end{equation}
where $\Rm$ is the magnetic Reynolds number, after which magnetic diffusion
destroys the field in the rotating region. We find a similar scaling of the
maximum magnetic field strength, $B_\textrm{rms, max}$ with the dimensionless
quantity
\begin{equation}
\Rmt = \frac{u_0 l_0}{u_rd_0},
\end{equation}
which we identify as the effective magnetic Reynolds number. Here $u_0$ and
$l_0$ are typical velocity and length scales respectively, $d_0$ is the
reconnection length, and $u_r$ is the characteristic reconnection speed. In
the case of the Gaussian vortex this is taken as the relative velocity of the
approaching flux tube. Figure \ref{flux_expuls_snaps} shows snapshots of
a typical simulation as it proceeds: the magnetic field after one winding
($t=5.0$), in a state close to the maximum field strength ($t=40.0$), as the reconnections start to drive the destruction of the field ($t=46.5$), and finally the quasi-steady state ($t=200.0$). The first plot in
Fig.~\ref{flux_expuls_analysis} shows the corresponding values of
$B_\mathrm{rms}$ versus time; the linear growth before the onset of
reconnections is apparent. The second plot in Fig.~\ref{flux_expuls_analysis}
shows the power-law relationship between $B_\mathrm{rms, max}$ and $\Rmt$: the
slope of the fit shown is $0.54 \pm 0.21$, in a reasonable agreement with
Eq.~(\ref{Moffatt_flux}).
One final test, results not presented here, was to ensure that no dynamo could be supported 
by driving the flux ropes with a two dimensional flow, i.~e.~ $\mathbf{u}=(u_x,u_y,0)$.

\begin{figure*}
   \begin{center}
      \psfrag{x}{x}
      \includegraphics[width=0.5\textwidth]{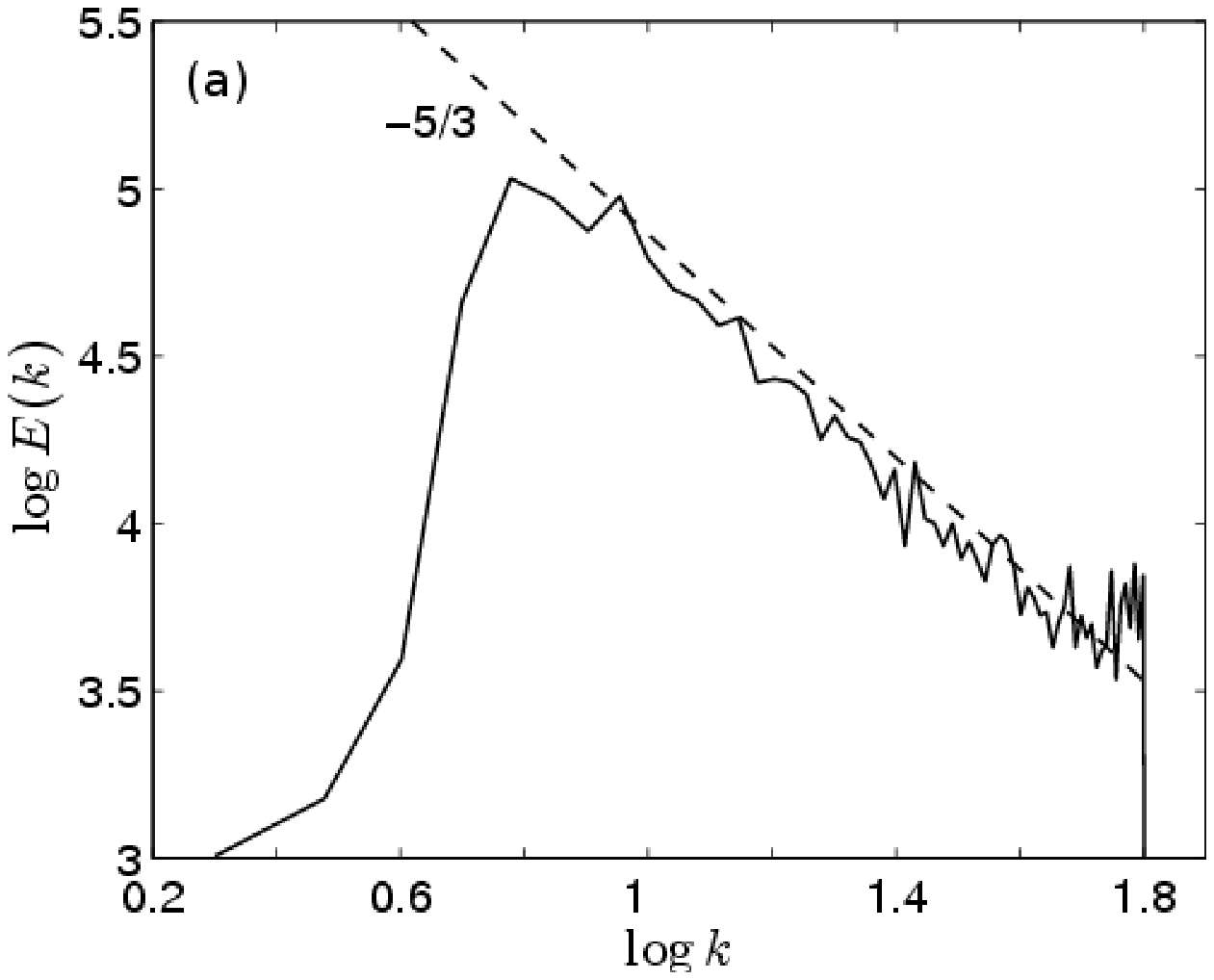}
    \qquad
      \includegraphics[width=0.4\textwidth]{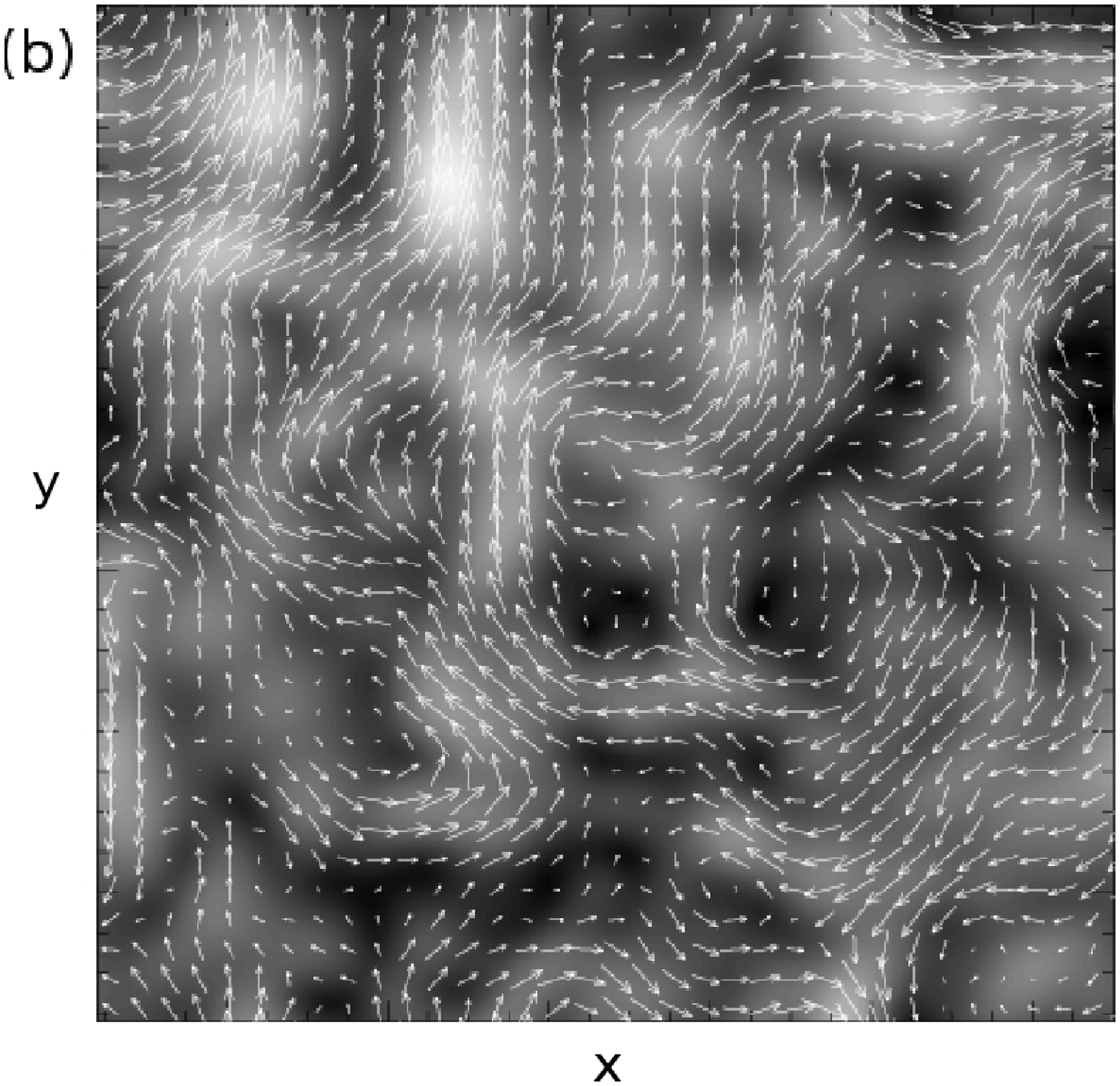}
\end{center}
      \caption{\label{KS_spectrum}
     {\bf(a)}  The energy spectrum $E(k)$
               as obtained by Fourier transform of Eq.~(\ref{uF}) with $N=20$,
               $k_1=10$ and $k_N=400$. The dashed line has $E(k) \propto k^{-5/3}$.
     {\bf(b)}  Slice in the $(x,y)$-plane of the vorticity field from {\bf(a)},
lighter shades indicating higher vorticity. Velocity vectors are shown in white.}
\end{figure*}

\section{Model of a turbulent flow}
Our next step is to choose the velocity field which drives the dynamo.
Following previous work \citep{Wilkin:2007}, to bypass the computational limitations of
direct numerical simulations (DNS), we use the so-called Kinematic Simulation
(KS) model. The KS model has primarily been used as a Lagrangian model of
turbulence and results are in good agreement with DNS
\citep{Fung:1998,Malik:1999,Osborne:2006}. This flow is known to be a
hydromagnetic dynamo \citep{Wilkin:2007}. The KS model prescribes the flow
velocity at a position $\mathbf{x}$ and time $t$ through the summation of
Fourier modes with randomly chosen parameters (which are then kept fixed),
according to \citep{Osborne:2006}:
\begin{equation}\label{uF}
{\bf u}({\bf x},t)= \sum_{n=1}^{N}\left(\mathbf{A}_n \times {\bf k}_n
\cos\phi_n + {\bf B}_n \times {\bf k}_n \sin\phi_n \right),
\end{equation}
where $\phi_n=\mathbf{k}_n \cdot {\bf{x}} + \omega_n t$, $N$ is the number of
modes, $\mathbf{k}_n$ and $\omega_n=k_n u_n$ are their wave vectors and
frequencies  (see \citep{Wilkin:2007} for details). The unit vectors
$\hat{\mathbf{k}}_n$ are chosen randomly, and
$\mathbf{k}_n=k_n\hat{\mathbf{k}}_n$ where $k_n$ is the wave number of the
$n$'th mode. We choose the directions of $\mathbf{A}_n$, and $\mathbf{B}_n$
randomly, imposing only orthogonality with $\hat{\mathbf{k}}_n$, which gives
\begin{equation}
   |\mathbf{A}_n \times \hat{\mathbf{k}}_n|=A_n,
 \end{equation}
and likewise for $\mathbf{B}_n$. We then select a kinetic energy spectrum $E(k)$ and set
  \begin{equation}
    A_n=B_n=\left[\tfrac{2}{3}E(k_n)\Delta k_n\right]^{1/2}.
  \end{equation}
This ensures that
  \begin{equation}
    \dfrac{1}{V}\int_V\tfrac{1}{2}|\mathbf{u}|^2\,dV=\int_0^{\infty} E(k) \, dk
\approx \sum_{n=1}^{N_k} E(k_n) \Delta k_n.
  \end{equation}
One of the main advantages of the KS
model is that we have complete
control of the energy spectrum, $E(k_n)$
via appropriate choice of $\mathbf{A}_n$ and $\mathbf{B}_n$.
We also note that $\nabla\cdot\mathbf{u}\equiv0$ by construction.
We adopt a modification of the von K\'{a}rm\'{a}n spectrum,
\begin{equation}\label{Ek}
   E(k)=k^4(1+k^2)^{-(2+p/2)}e^{-\frac{1}{2}(k/k_{N})^2},
\end{equation}
which reduces to $E(k)\propto k^{-p}$ in the inertial range, $1\ll k\ll k_N$,
with $k=1$ at the integral scale; $p=5/3$ produces the Kolmogorov spectrum,
and $k_N$ is the cut-off scale. Figure \ref{KS_spectrum} shows the energy
spectrum of the KS flow, obtained numerically after fast Fourier transforming
$\mathbf{u}$ calculated from Eq.~(\ref{uF}) on a $128^3$ mesh. We also show a
slice, in the $(x,y)$-plane, of the corresponding vorticity field.

The results presented below have been obtained with $k_1=2\pi$ and
$k_N=16\pi$, so that the smallest velocity scale is $l_N=2\pi/k_N=0.125$. For
comparison, the reconnection scale is adopted as $d_0=l_N/4$ unless stated
otherwise. With this prescription, the effective magnetic Prandtl number in
our model is larger than unity.

\begin{figure}
  \begin{center}
    \includegraphics[width=0.4\textwidth]{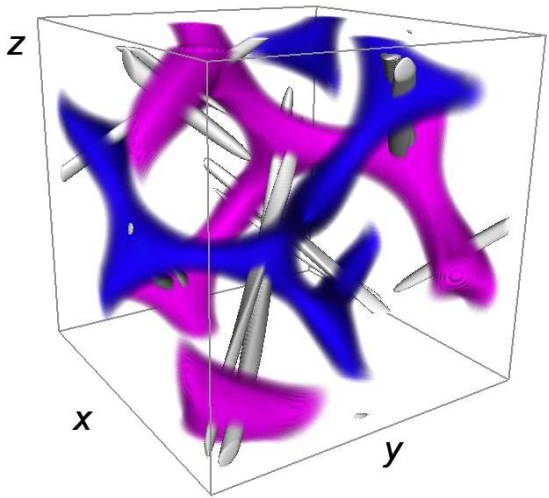}
    \caption{\label{DVR_ABC}(Colour online) The distribution of $|\mathbf{u}|$ in
the 111 ABC flow Eq.~(\ref{ABC_flow}) is shown in colour, with the
magnetic isosurfaces with $|\mathbf{B}|=3.5B_{\textrm{rms}}$, obtained by
solving the induction equation, shown in grey scale. Each point on a $3D$ mesh
is assigned an opacity and colour, depending on $|\mathbf{u}|$. Regions with
$|\mathbf{u}|>2.5u_{\textrm{rms}}$ are coloured purple and
those where $|\mathbf{u}| \approx 0$ are coloured blue.}
\end{center}
\end{figure}

To check if our results depend on the form of the flow, we
also use an ABC flow of the form \citep{Childress:1995},
\begin{equation}\label{ABC_flow}
\mathbf{u}=(\cos ky+\sin kz,\sin kx+\cos kz, \cos kx+\sin ky)\;,
\end{equation}
known as the 111 ABC flow.
Dynamo action driven by ABC flows has been studied extensively
\citep{Galloway:1984}. This particular flow has eight stagnation points for
$0<k_x<2\pi$. In the plane in which the flow converges to a particular
stagnation point, the magnetic field is advected, and becomes elongated in the
direction of the streamlines which diverge from the stagnation point. The
resulting magnetic structures are commonly described as `magnetic flux cigars'
\citep{Dorch:2000}. Figure~\ref{DVR_ABC} shows such magnetic structures
produced by our numerical solution of the induction equation, along with the
corresponding velocity field.

\section{Diffusive and reconnection-based dynamos}\label{DiffvsRecon}
Comparisons of the solution of the induction equation with those produced by
the flux rope model are not straightforward because of the difference in the
control parameters of the two models: the magnetic Reynolds number
$R_\textrm{m}=u_0 l_0/\eta$ and the reconnection length $d_0$, respectively.
Following the previous section, we introduce the effective magnetic Reynolds
number as $\tilde{R}_\textrm{m}=u_0 l_0/(u_\textrm{r}d_0)$, where
$u_\textrm{r}$ is the characteristic reconnection speed.

We find that the dynamo based on
reconnections is  more efficient than the diffusion-based dynamo, in the sense
that the growth rate of magnetic field of the former is significantly larger
when $R_\textrm{m}\approx\tilde{R}_\textrm{m}$. Therefore, in order to achieve
conservative conclusions, we compare dynamos with \textit{similar growth
rates\/} of magnetic field. Thus, we have $R_\textrm{m}>\tilde{R}_\textrm{m}$
in the models which we compare.

\begin{figure}
  \begin{center}
    \includegraphics[width=0.4\textwidth]{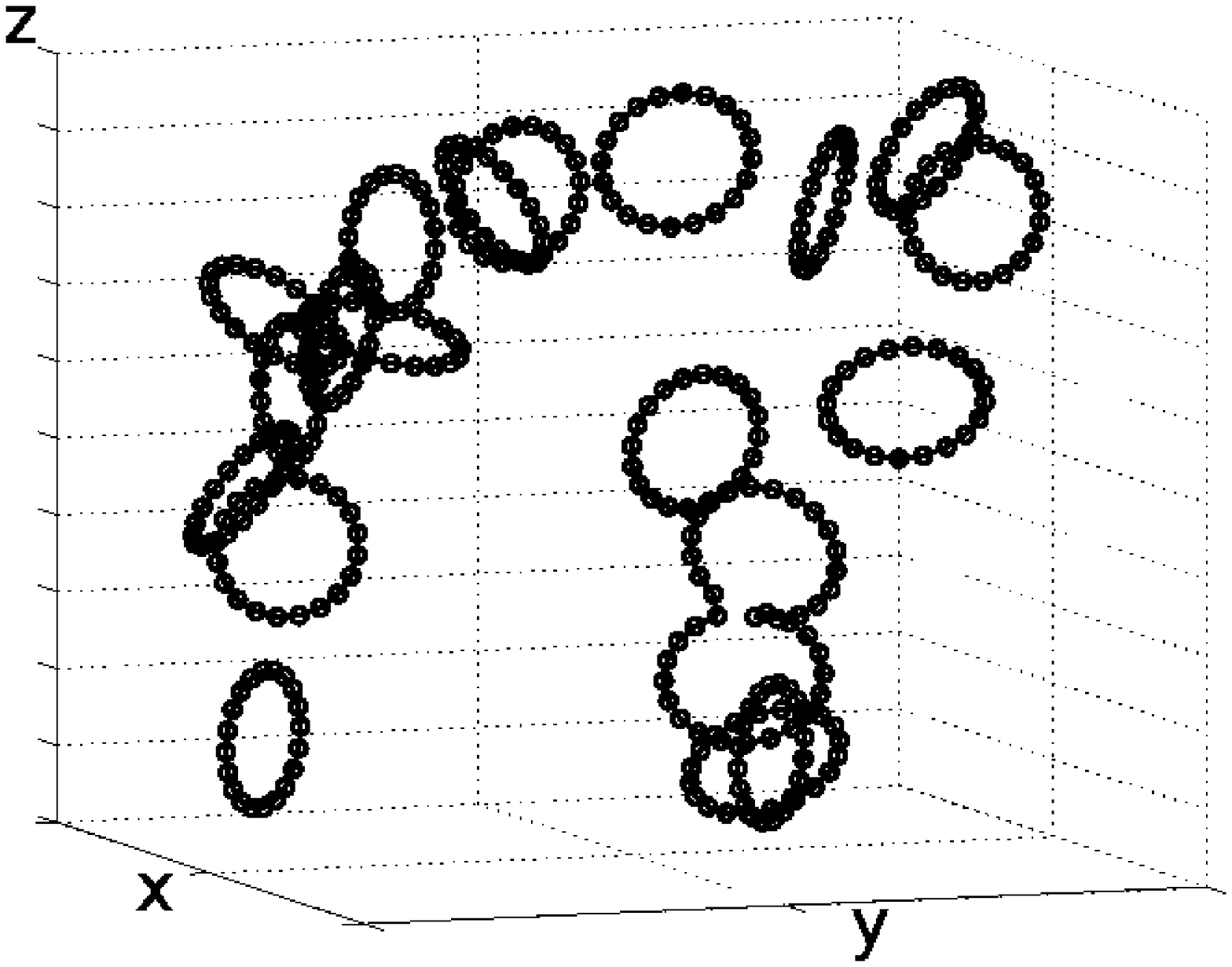}
    \includegraphics[width=0.4\textwidth]{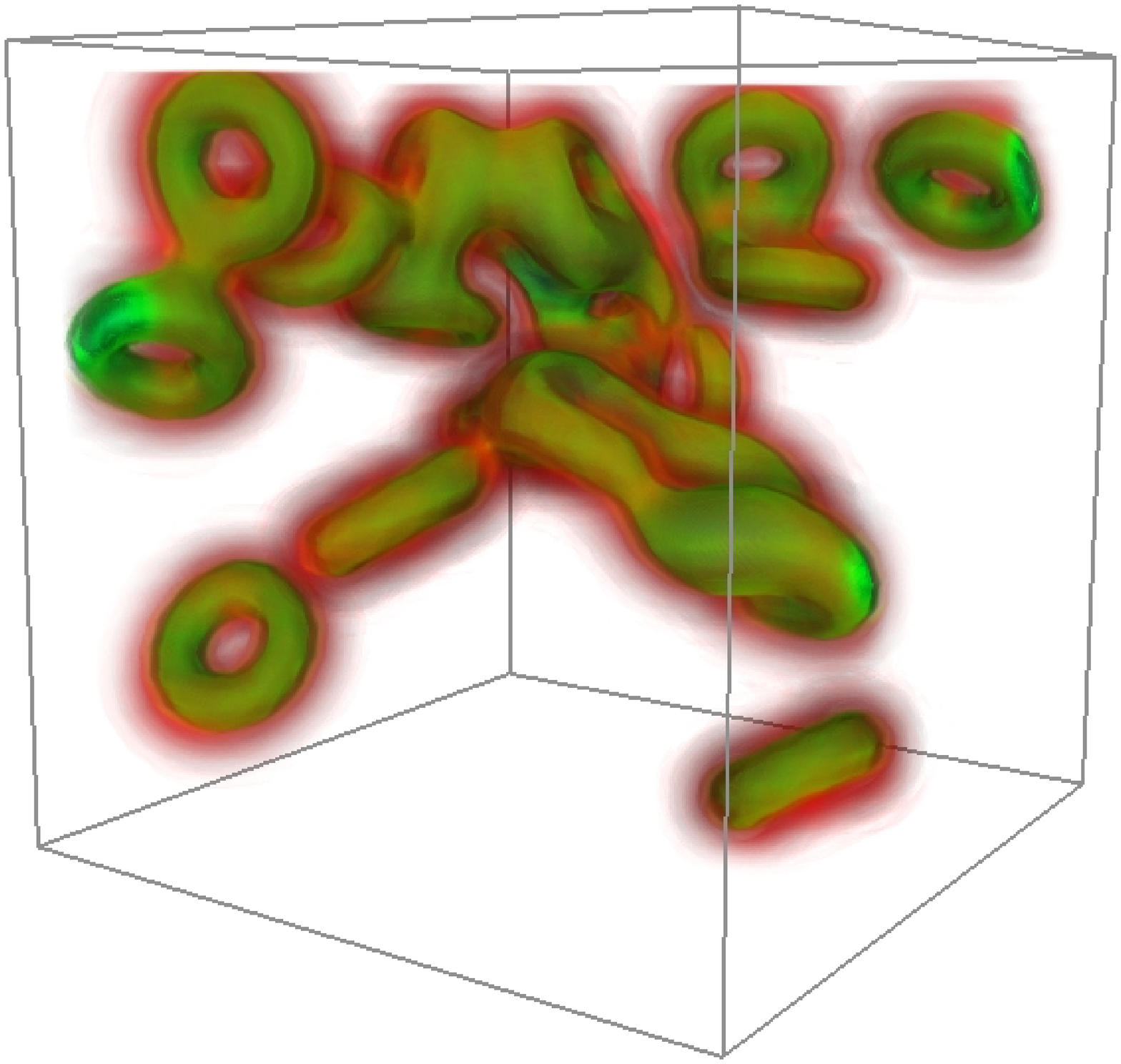}
    \caption{\label{initial_setup}(Colour online) Initial conditions for the magnetic
field in the ABC and KS simulations. The upper panel shows the initial
set of flux ropes populated by the trace particles. The lower panel presents a
direct volume rendering of a similar set of closed flux ropes, but now smoothed
using a Gaussian kernel (\ref{Gauss_smooth}). The smoothed magnetic field is
used to initialise the induction equation simulations.}
\end{center}
\end{figure}

An advantageous property of both KS and ABC flows is their analytic nature,
which means that we can follow (Lagrange-like) fluid particles in the flow without using an Eularian mesh.
The initial condition of our simulations is an ensemble of
random closed magnetic loops; both the induction equation and the flux rope
model are evolved with the same velocity field (apart from the overall
normalisation to provide comparable growth rates of magnetic field). The
initial condition for the induction equation is obtained by Gaussian smoothing
of the magnetic field in the ropes, where we define the smoothed field
$\tilde{\mathbf{B}}(\mathbf{x})$, as
\begin{equation}\label{Gauss_smooth}
\mathbf{\tilde{\mathbf{B}}}(\mathbf{x})
=\frac{1}{2\pi \sigma^2}
\int_V e^{-|\mathbf{x}-\mathbf{y}|^2/2 \sigma^2}\mathbf{B}(\mathbf{y})
\,d\mathbf{y}^3.
\end{equation}
Importantly, this procedure preserves the solenoidality of the field, i.~e.,
$\nabla\cdot\tilde{\mathbf{B}}=0$. Figure ~\ref{initial_setup} shows the
smoothed initial magnetic field used in a typical simulation, along with the
corresponding flux rope setup.

The induction equation is solved using the Pencil Code \citep{PC:2002}, which
implements a high-order finite-difference scheme, on a $256^3$ mesh with
$1000<\Rm<1500$ in a periodic box. Simulations with the KS velocity field had
$k_1=2\pi$, $k_N=16\pi$, and $p=-5/3$; here $\Rm$ is based on the
 largest velocity  scale $2\pi/k_1$.
In a separate simulation flux ropes are advected and stretched by the {\it
same\/} velocity field, where the positions of the trace particles are evolved
using a $4^{\textrm{th}}$ order Runge--Kutta scheme, with a time step of
$l_N/(20u_N)$. The algorithm for inserting and removing particles is applied
every time step, and the reconnection algorithm, every ten time steps. We
choose $d$ to be 1/4 of the smallest length scale in the flow and set
$d_0/d=1.5$, where $d_0$ is the reconnection length scale.

\begin{figure}
  \begin{center}
    \psfrag{x}{x}
    \psfrag{y}{y}
    \psfrag{z}{z}
    \psfrag{t1}{t=0.09}
    \psfrag{t2}{t=0.5}
    \psfrag{t3}{t=1.0}
    \psfrag{t4}{t=1.5}
    \includegraphics[width=0.4\textwidth]{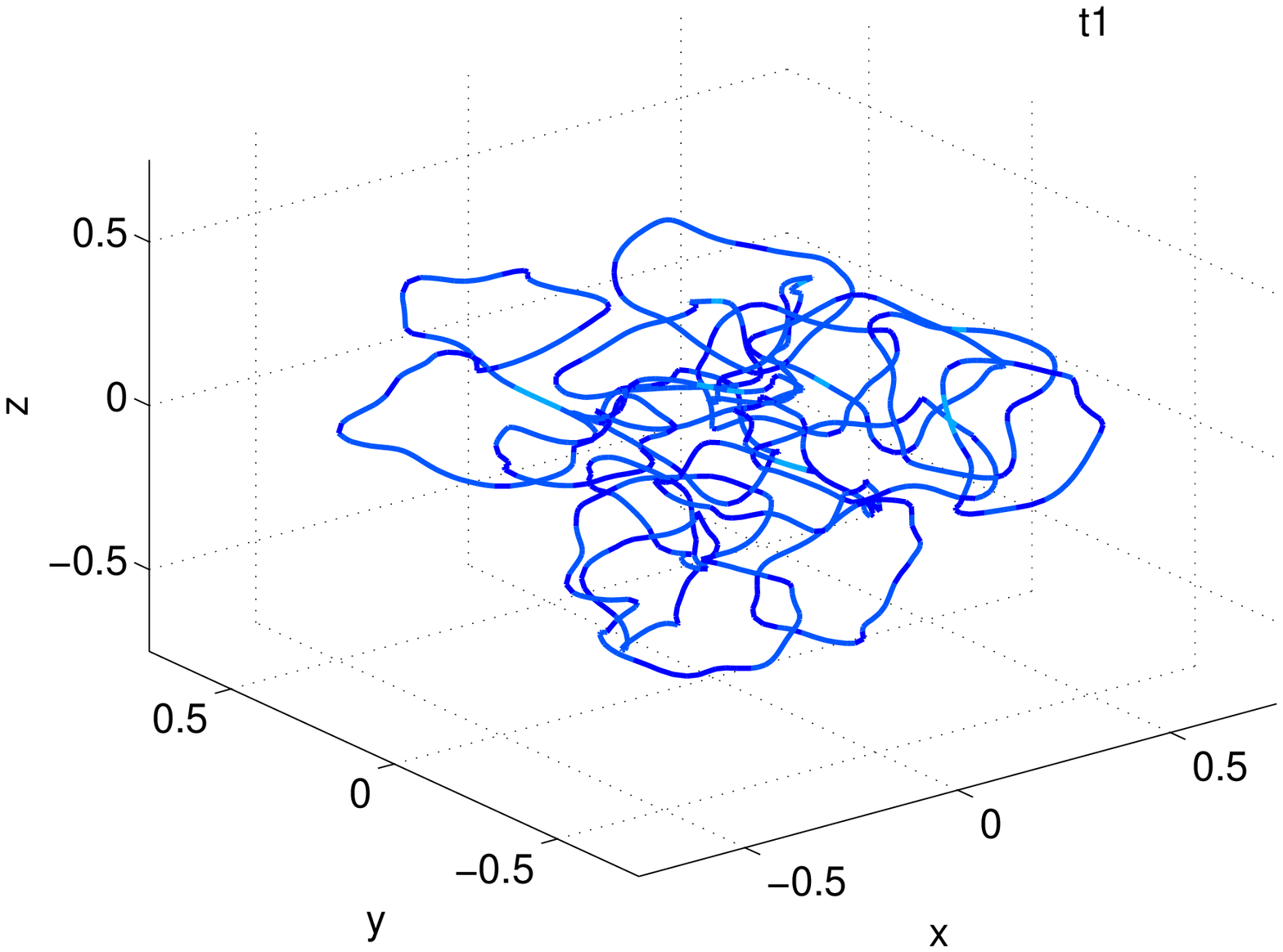}
    \includegraphics[width=0.4\textwidth]{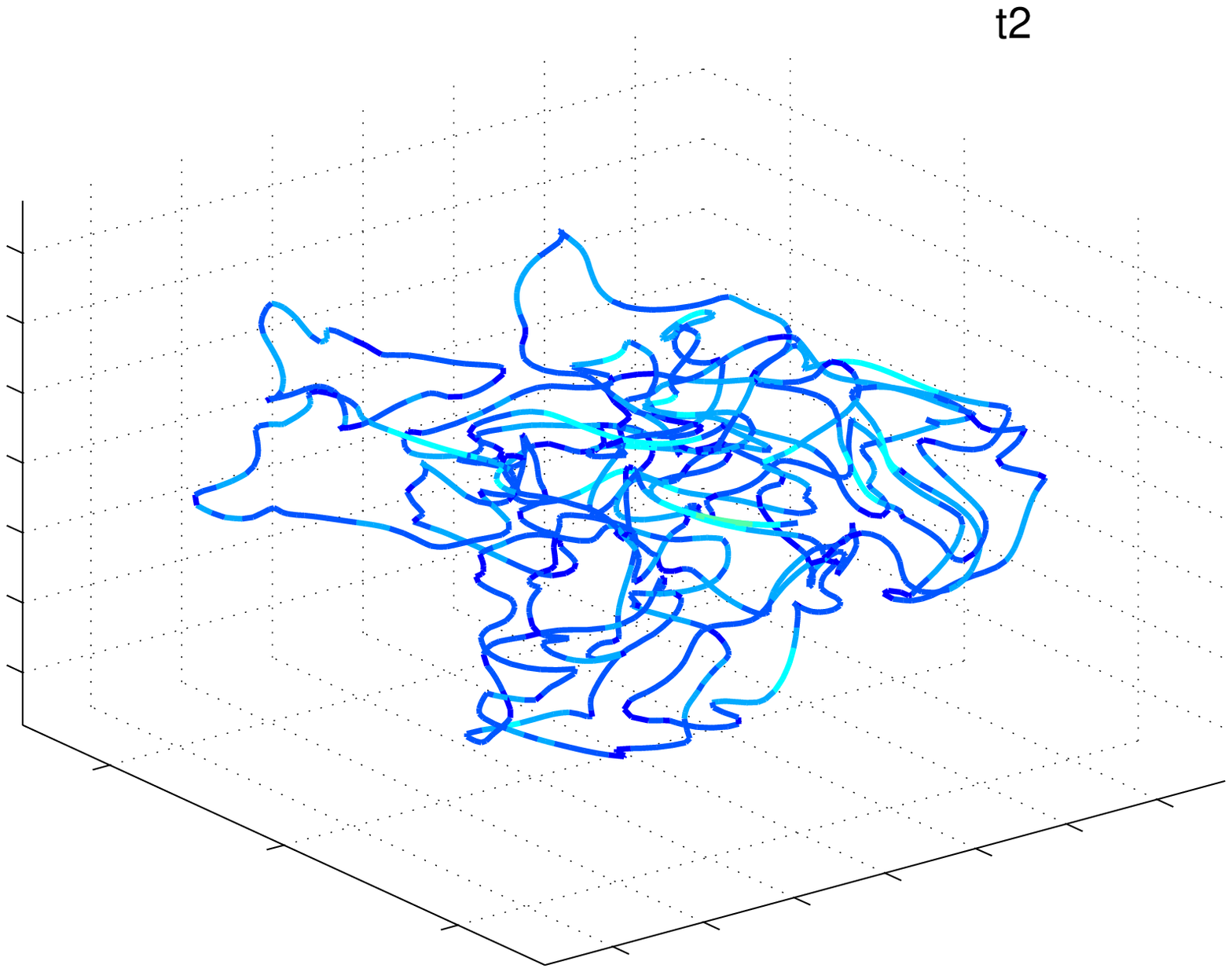}
    \includegraphics[width=0.4\textwidth]{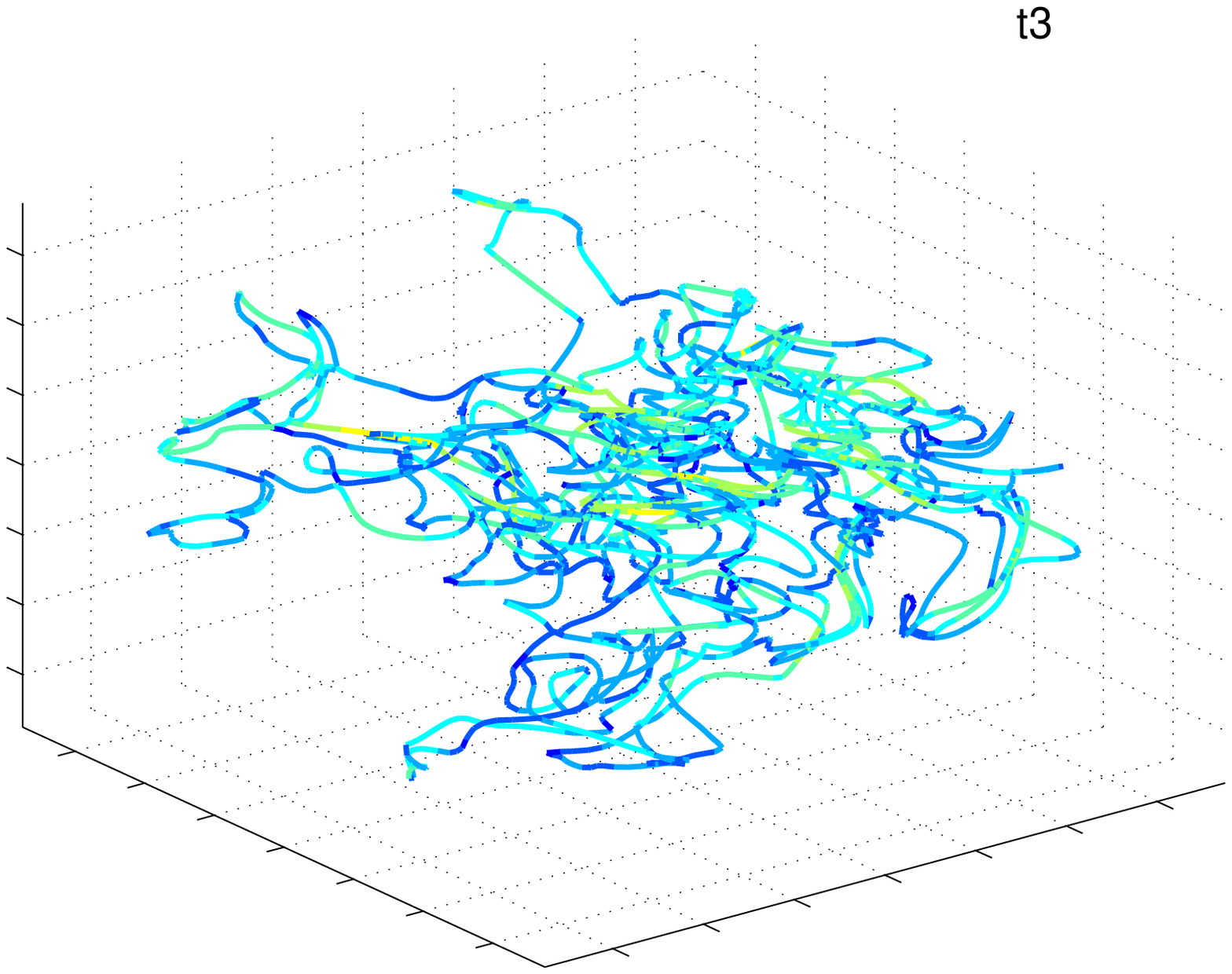}
    \includegraphics[width=0.4\textwidth]{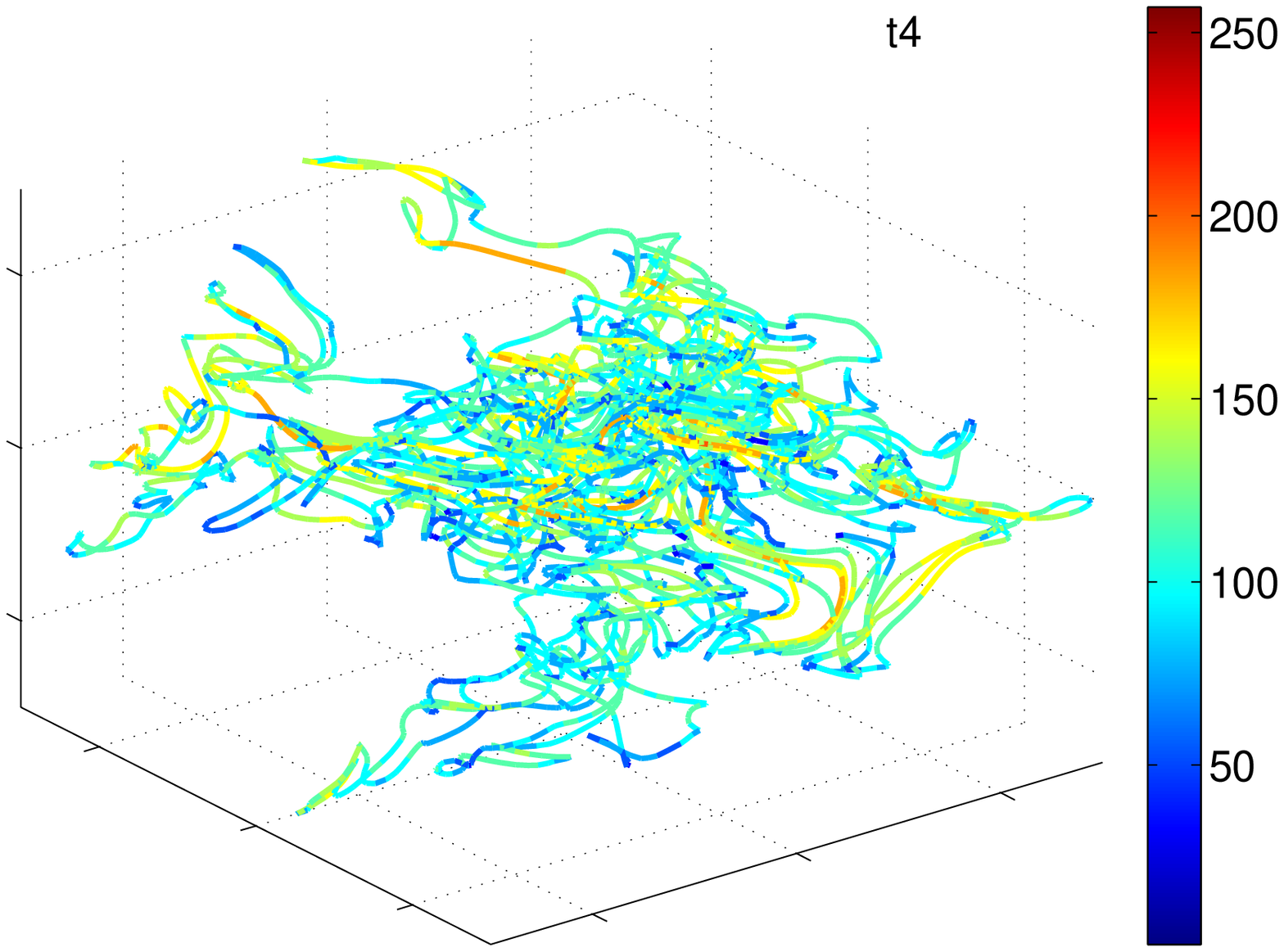}
    \caption{\label{KS_snap}(Colour online) The evolution of magnetic flux
tubes in the KS flow: snapshots taken at times given in the upper right corner of each panel.
Magnetic field strength is colour coded, with the colour bar shown next to the
last snapshot. Note the overall increase of magnetic field strength as time proceeds.}
\end{center}
\end{figure}

\begin{figure}
  \begin{center}
    \psfrag{x}{x}
    \psfrag{y}{y}
    \psfrag{z}{z}
    \psfrag{t1}{t=1.0}
    \psfrag{t2}{t=2.5}
    \psfrag{t3}{t=5.0}
    \psfrag{t4}{t=7.5}
    \includegraphics[width=0.4\textwidth]{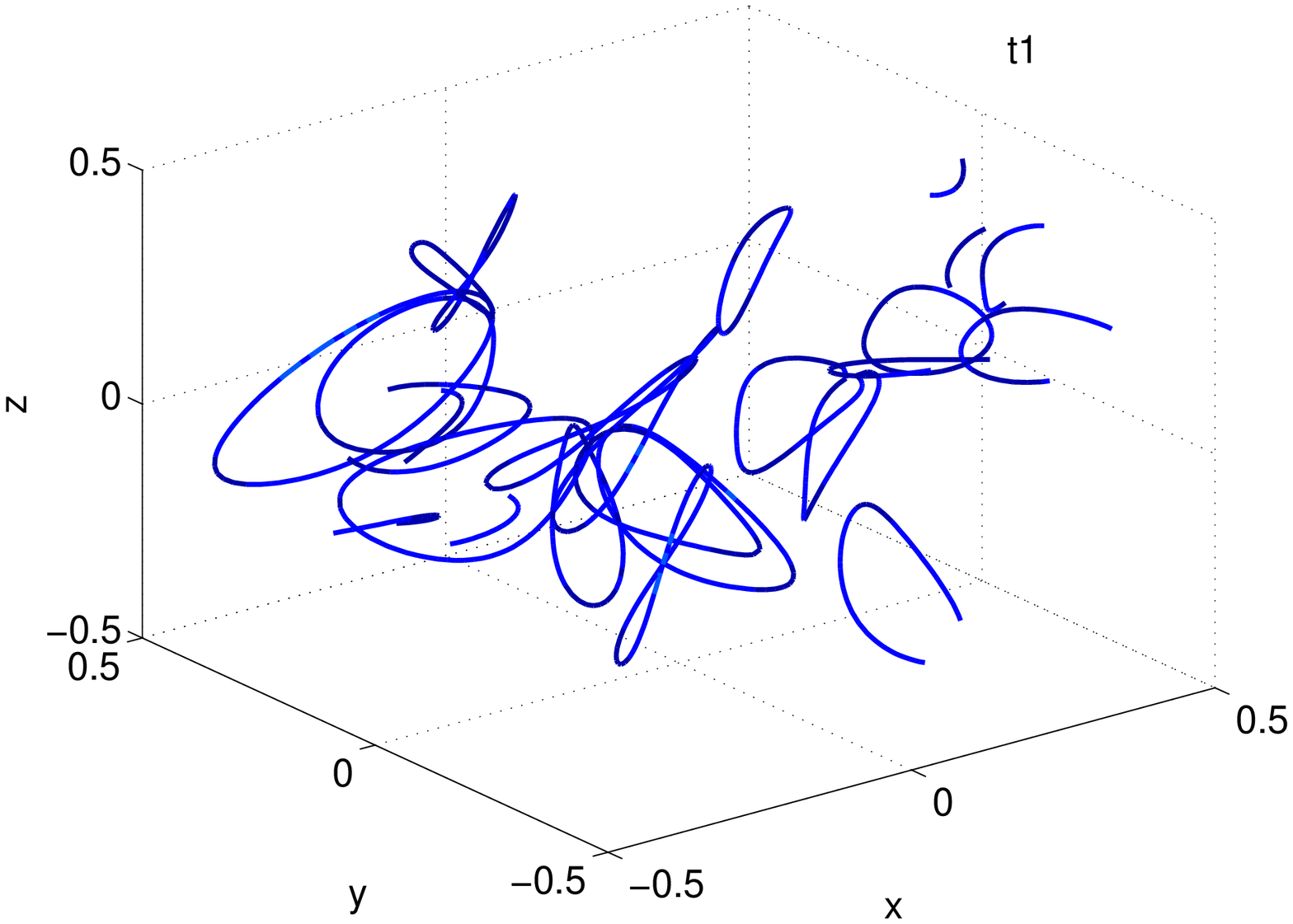}
    \includegraphics[width=0.4\textwidth]{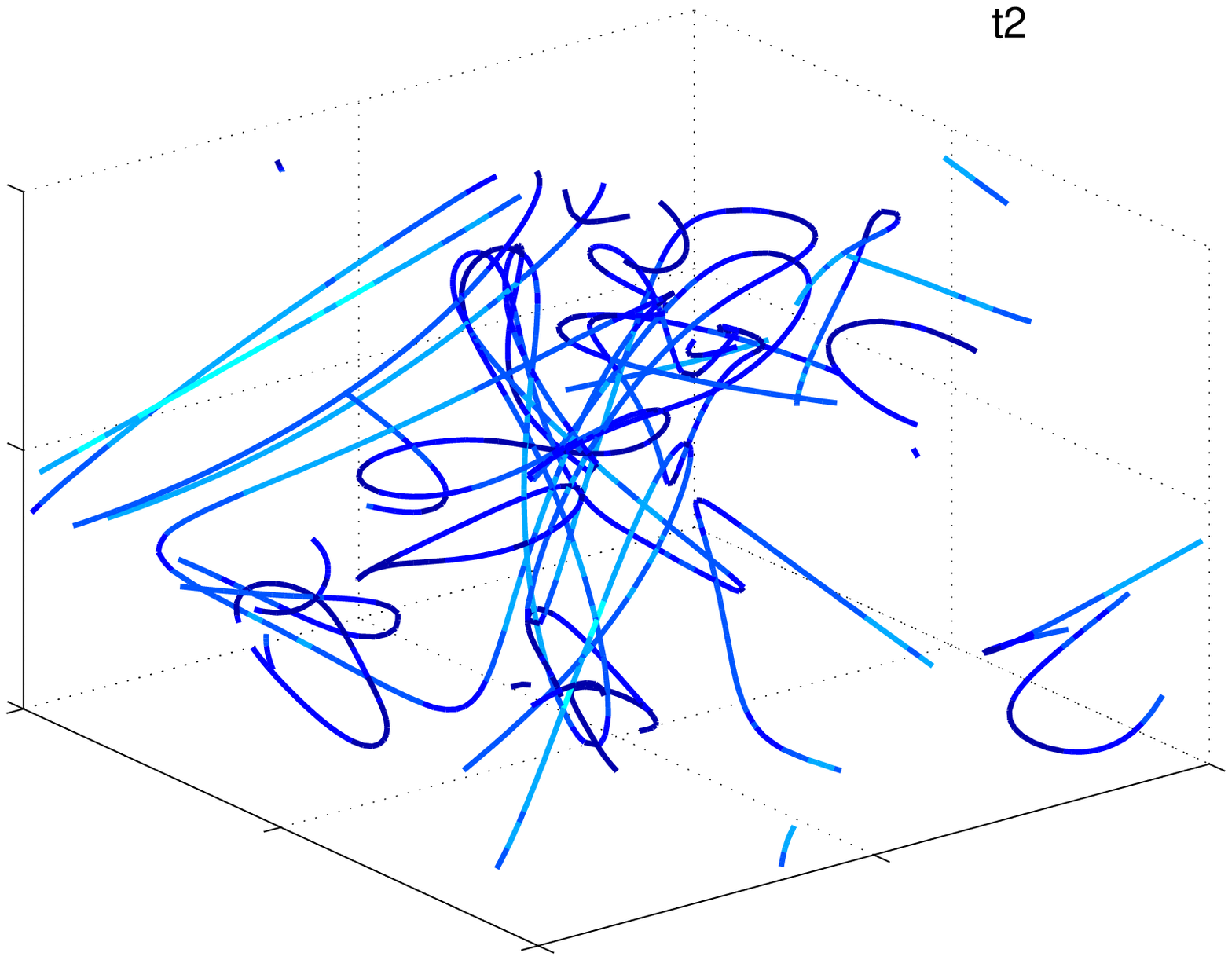}
    \includegraphics[width=0.4\textwidth]{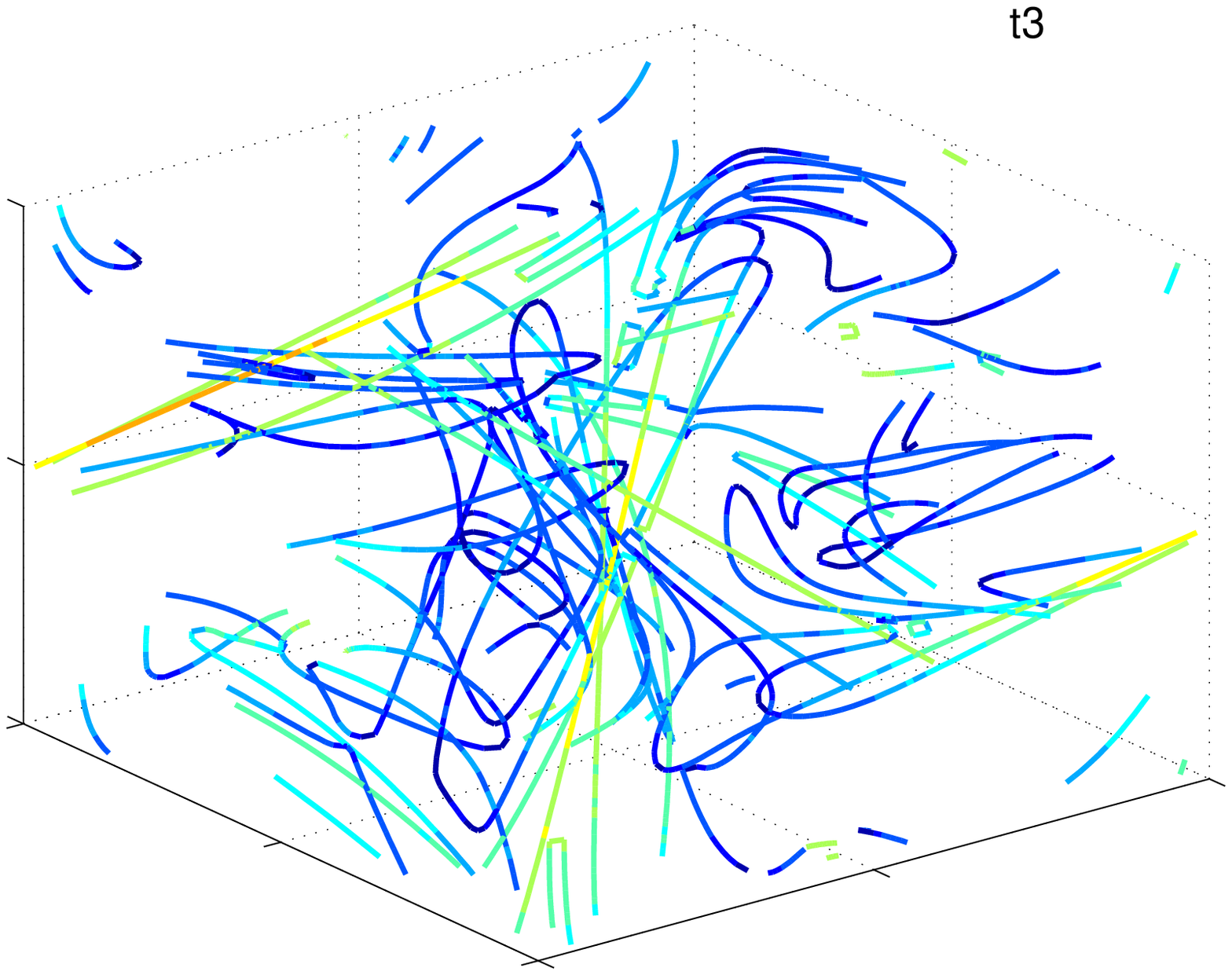}
    \includegraphics[width=0.4\textwidth]{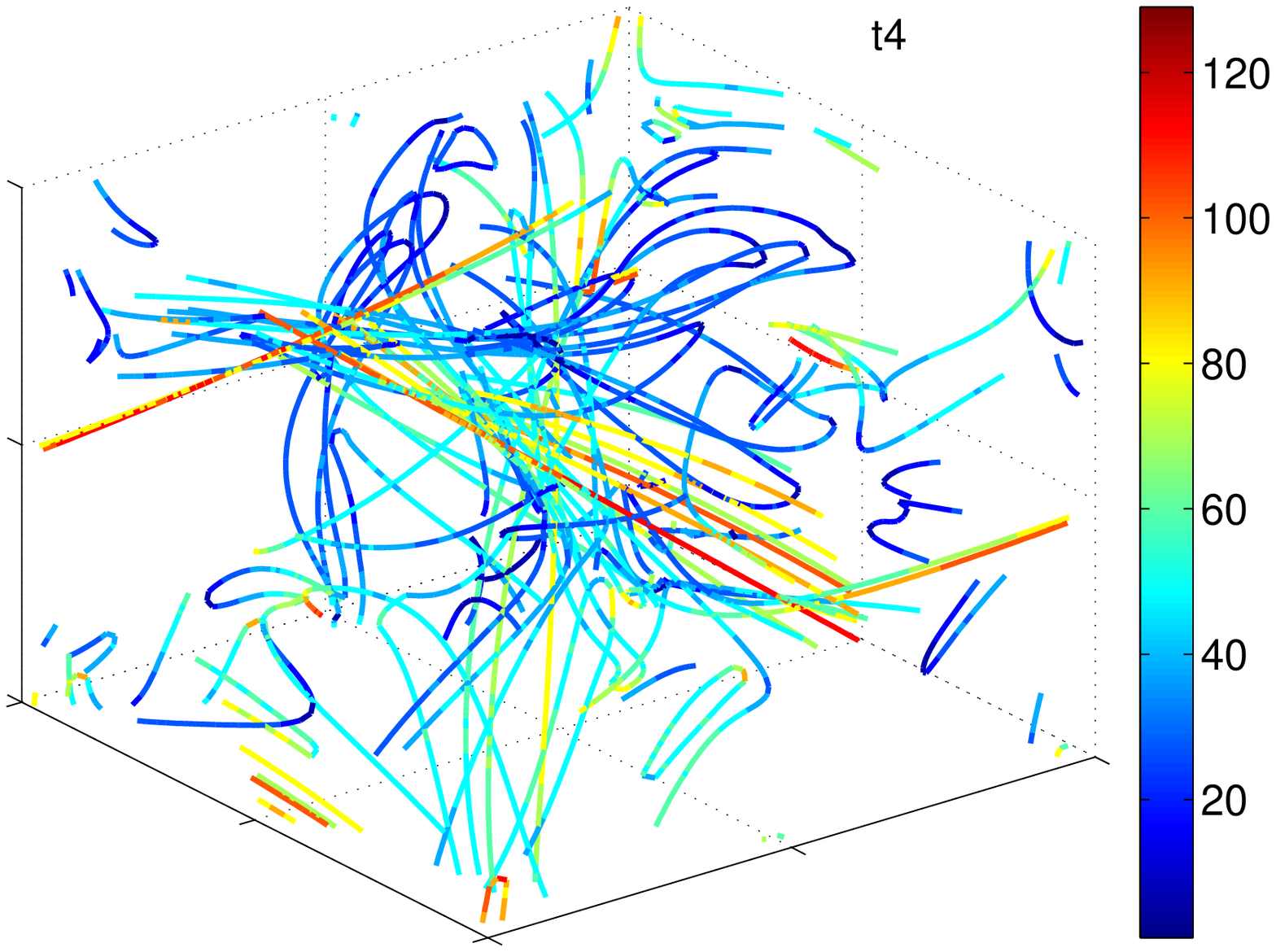}
    \caption{\label{ABC_snap}(Colour online) As in Fig.~\ref{KS_snap}, but for
the 111 ABC flow. Magnetic field has been stretched into `flux cigars' which are
    even more apparent when the field is later smoothed.}
\end{center}
\end{figure}

In Fig.~\ref{KS_snap} we present snapshots of the magnetic field from the flux
rope simulations, driven by the KS flow, as it evolves, Fig.~\ref{ABC_snap}
shows the corresponding evolution in the ABC flow. In the case of the ABC
 flow, one notices an anticorrelation between the curvature of the flux tube
 and the magnetic field strength (colour coded) as suggested by
 \citet{Sheck:2002}.

Our model is deliberately oversimplified with respect to the (incompletely
understood) physics of magnetic reconnections. Nevertheless, we can argue that
our model is conservative with respect to the reconnection efficiency. The
reconnecting segments of magnetic lines in our model approach each other at a
speed $u_\mathrm{r}\simeq u_0\Rey^{-1/4}$ for the Kolmogorov spectrum, equal
to velocity at the \textit{small\/} scale $d_0\ll l_0$ with $l_0$ the
energy-range scale of the flow and $d_0$ assumed to be close to the turbulent
cut-off scale. If the magnetic field is strong enough, the Alfv\'en speed
$V_\mathrm{A}$, which controls magnetic reconnection in more realistic models,
is of order $u(l_0)$. Then $u_\mathrm{r}\ll V_\mathrm{A}$ and our model is
likely to underestimate the efficiency of reconnections. The Sweet--Parker
reconnection proceeds at a speed of order $V_\mathrm{A}\Rm^{-1/2}$, whereas
the Petschek reconnection speed is comparable to $V_\mathrm{A}/\ln\Rm$
\citep{priest:2000}. For $u_0\simeq V_\mathrm{A}$ and $\Rm\simeq\Rey\gg1$, the
reconnection rate in our model is larger than the former but much smaller than
the latter.

\section{Coupling with the Navier--Stokes equation and dynamo saturation}\label{NLM}
A remarkable feature of the model of reconnecting
magnetic flux tubes suggested here is that is admits straightforward extension
to include the back-reaction of magnetic field on the flow via the Lorentz
force. Such a generalisation is presented in this section. To obtain a flow
similar to the KS flow (\ref{uF}) as a solution of the Navier--Stokes
equation, we include a driving force shown here after the viscous term:
\begin{equation}
\frac{D\mathbf{u}}{D t}=
-\frac{\nabla P}{\rho} + \frac{\mathbf{J} \times \mathbf{B}}{4\pi\rho}
+ \frac{1}{\Rey}\nabla^2 \mathbf{u} + 
\frac{\mathbf{u}_{\textrm{KS}}-\mathbf{u}}{\tau},
\end{equation}
where $D/Dt=\partial/\partial t+\mathbf{u}\cdot\nabla$ is the convective
(Lagrangian) derivative, $\Rey$ is the Reynolds number,
$\mathbf{u}_{\textrm{KS}}$ is the KS velocity field  (\ref{uF}), and $\tau$ is
a certain relaxation time. The smaller is $\tau$, the closer $\mathbf{u}$ is
to the KS flow. Since we assume that magnetic field is localised within flux
ropes, magnetic pressure must be balanced by some other force, presumably by
gas pressure, so we assume that $\nabla(P+B^2/8\pi)=0$, and only the magnetic
tension force $(\mathbf{B}\cdot\nabla)\mathbf{B}$ remains to be balanced in
the Navier--Stokes equation. Neglecting viscosity, $\Rey\to\infty$,
we then obtain
\begin{equation}\label{NSt}
\frac{D\mathbf{u}}{D t}=
\frac{1}{8\pi}(\mathbf{B}\cdot\nabla)\mathbf{B}+
\frac{\mathbf{u}_{\textrm{KS}}-\mathbf{u}}{\tau}.
\end{equation}
If magnetic field is confined into thin ropes and aligned with their axes,
magnetic tension involves the directional derivative of magnetic field along
the rope axis alone,
$(\mathbf{B}\cdot\nabla)\mathbf{B}=B\,\partial\mathbf{B}/\partial s$, where
$s$ is the distance measured along the rope. Thus, it is sufficient to have
magnetic field defined on magnetic loops (rather than at any position in the
volume) in order to calculate magnetic tension force.

We require a solution of
Eq.~(\ref{NSt}) at the changing positions of the trace particles, so we need,
essentially, a Lagrangian solution of this equation. Assuming that the flow is
close to the relaxed state and does not change rapidly, we put $D\mathbf{u}/D
t\approx0$ to obtain the trace particle velocities as
\begin{equation}\label{reducedmodel}
\mathbf{u}\approx\mathbf{u}_{\textrm{KS}}
        +\tau B\frac{\partial\mathbf{B}}{\partial s}.
\end{equation}
This approximation filters out rapid wave motions, e.g., Al\-fv\'en waves,
which simplifies numerical simulations. We use this
approximation to study the saturation of the dynamo action in
Section~\ref{DM}, where we consider rather long time intervals.

On the other hand, our model also allows us to include Al\-fv\'en
waves and their nonlinear interactions. For this purpose we assume
that $|D\mathbf{u}/Dt|\gg|\mathbf{u}_{\textrm{KS}}-\mathbf{u}|/\tau$ and the
Navier--Stokes (or rather Euler) equation reduces to
\begin{equation}\label{fullmodelnoks}
\frac{D\mathbf{u}}{D t}\approx
B \frac{\partial \mathbf{B}}{\partial s},
\end{equation}
which, can be coupled with the equation for a frozen-in magnetic field
$D\mathbf{B}/Dt=(\mathbf{B}\cdot\nabla)\mathbf{u}$, written in a similar
form:
\begin{equation}\label{bvolution}
\frac{\partial \mathbf{B}}{\partial t}
=B\frac{\partial \mathbf{u}}{\partial s}.
\end{equation}
Imposing a homogeneous magnetic field $\mathbf{B}_0$ and
a weak perturbation, and linearising these equations leads to the wave equation
describing the Alfv\'en waves. Since we assume that magnetic pressure is
precisely balanced by gas pressure, our model does not admit compressible
waves.

The nonlinearity requires that we make two changes to our
numerical calculations. Firstly in our model
$B\,\partial \mathbf{B}/ \partial s$ is only defined at positions on the
magnetic line (flux tube), and so the velocity field can only be evolved at
those positions. The fourth-order Runge--Kutta time stepping scheme used in
the kinematic regime is not suitable as it requires velocity field at
positions where magnetic field is not defined. Therefore, we use the
three-step Adams--Bashforth scheme instead to evolve the positions of the
trace particles:
\begin{eqnarray*}\label{adamb}
  \mathbf{x}_{n+1} &=&
\mathbf{x}_n+\frac{h}{12}(23\mathbf{u}_n-16\mathbf{u}_{n-1}+5\mathbf{u}_{n-2}), \\
  t_{n+1} &=& t+h, \nonumber
\end{eqnarray*}
where $h$ is the size of the timestep.

The differentiation of the magnetic field along the flux tubes requires an
improved accuracy for the positions of newly introduced trace particles in a
stretched flux tube. A first-order prescription (\ref{lin_interp}) is no longer
accurate enough and we replace it by a second-order interpolation scheme.
Consider a section of magnetic line traced by three particles at positions
$\mathbf{x}_1$, $\mathbf{x}_2$ and $\mathbf{x}_3$. If the distance between
$\mathbf{x}_2$ and $\mathbf{x}_3$ becomes greater than $d$, we introduce a new
particle at the position $\mathbf{x}_4$ given by
\begin{eqnarray*}
\mathbf{x}_4
&=&\mathbf{x}_1-[(\mathbf{x}_3-\mathbf{x}_1)-4(\mathbf{x}_2-\mathbf{x}_1)]\mu\\
&&\mbox{}+[2(\mathbf{x}_3-\mathbf{x}_1)-4(\mathbf{x}_2-\mathbf{x}_1)]\mu^2,
\end{eqnarray*}
where $\mu$ is a parameter. For $\mu=0.75$, the new particle
is placed between $\mathbf{x}_2$ and $\mathbf{x}_3$ as required. In tests,
in particular with Alfv\'en waves, we found a substantial improvement in the
accuracy of the solution with this higher-order scheme. However kinematic
results show no quantifiable difference between the two schemes.

\begin{figure}
  \begin{center}
    \psfrag{1}{\large$\mathbf{r}_{i+1}$}
    \psfrag{2}{\large$\mathbf{r}_{i}$}
    \psfrag{3}{\large$\mathbf{r}_{i-1}$}
    \psfrag{d}{\large$\ell_i$}
    \psfrag{e}{\large$\ell_{i-1}$}
    \includegraphics[height=0.4\textwidth,angle=270]{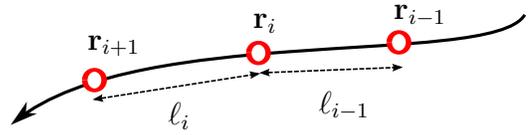}
    \caption{\label{derive_scheme}A sketch of tracer
particles that populate the flux tubes illustrating Eqs.~(\ref{first_deriv})
and (\ref{second_deriv}.)}
  \end{center}
\end{figure}

The directional derivative of magnetic field has to be calculated carefully in
our case since the separation of trace particles is not constant. We use the
following numerical schemes to evaluate the first derivative:
\begin{equation}\label{first_deriv}
  \mathbf{B}_i'=\frac{\ell_{i-1} \mathbf{B}_{i+1}
    +(\ell_i-\ell_{i-1})\mathbf{B}_i-\ell_i\mathbf{B}_{i-1}}{2\ell_i\ell_{i-1}}
    +O(\ell^2),
\end{equation}
where
the
notation is defined in Fig.~\ref{derive_scheme}, with $\mathbf{r}$
replacing
$\mathbf{B}$. For $\ell_i=\ell_{i-1}=h$, we recover a commonly
used finite difference scheme.

\subsection{Saturated dynamos}\label{DM}
Our starting point here is Eq.~(\ref{reducedmodel}) for the velocity field.
At each position on a flux tube, $\mathbf{x}^{(i)}$, we calculate the KS
velocity field using Eq.~(\ref{uF}), and modify it with magnetic tension
force. The directional derivative of magnetic field along the
tube, $\partial \mathbf{B}/ \partial s$, is computed using
Eq.~(\ref{first_deriv}).

The details of the simulations are similar to those in the kinematic regime.
We find our timestep of $l_N/(20u_N)$ to be sufficient to capture the dynamics
of the motion (we found no noticeable difference between simulations with the
timestep set an order of magnitude smaller than this). The relaxation time
$\tau$ is set to be the same as the timestep, $\tau=O(10^{-5})$.

\begin{figure}
  \begin{center}
    \psfrag{x}{$t/t_0$}
    \psfrag{y}{$\log B_\mathrm{rms}$}
    \includegraphics[width=0.4\textwidth]{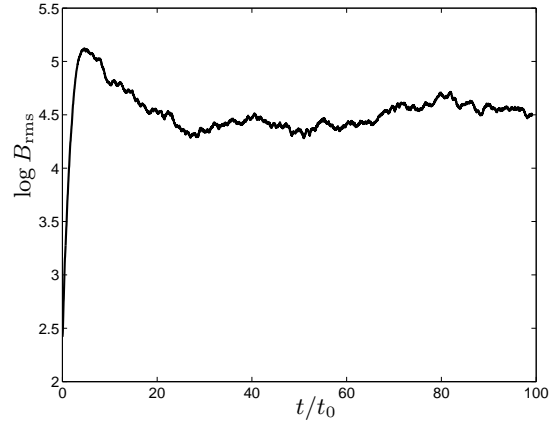}
    \caption{\label{saturated_brms}
The evolution of the root-mean-square
(r.m.s.) magnetic field  $B_\mathrm{rms}$ in nonlinear dynamo driven by the
KS flow (\ref{uF}) with the nonlinearity(\ref{reducedmodel}). Exponential
growth at $t\lesssim4$ is followed by a saturated state where magnetic energy
density fluctuates around a roughly constant level. The unit time is the
kinematic time scale at the largest scale in the flow, $t_0=l_0/u_0$.
}
\end{center}
\end{figure}

Figure~\ref{saturated_brms} shows the r.m.s.\ magnetic field strength
$B_\mathrm{rms}$ as a function of time, where the initial exponential growth
is followed, at $t\gtrsim4$, by the saturation of the dynamo action, with
$B_\mathrm{rms}$ fluctuating around a roughly constant level.

\begin{figure*}
  \begin{center}
    \psfrag{x}{x}
    \psfrag{y}{y}
    \psfrag{z}{z}
    \psfrag{t1}{t=0.6}
    \psfrag{t2}{t=1.9}
    \psfrag{t3}{t=3.9}
    \psfrag{t4}{t=25.0}
    \psfrag{t5}{t=50.0}
    \psfrag{t6}{t=85.0}
    \includegraphics[width=0.457\textwidth]{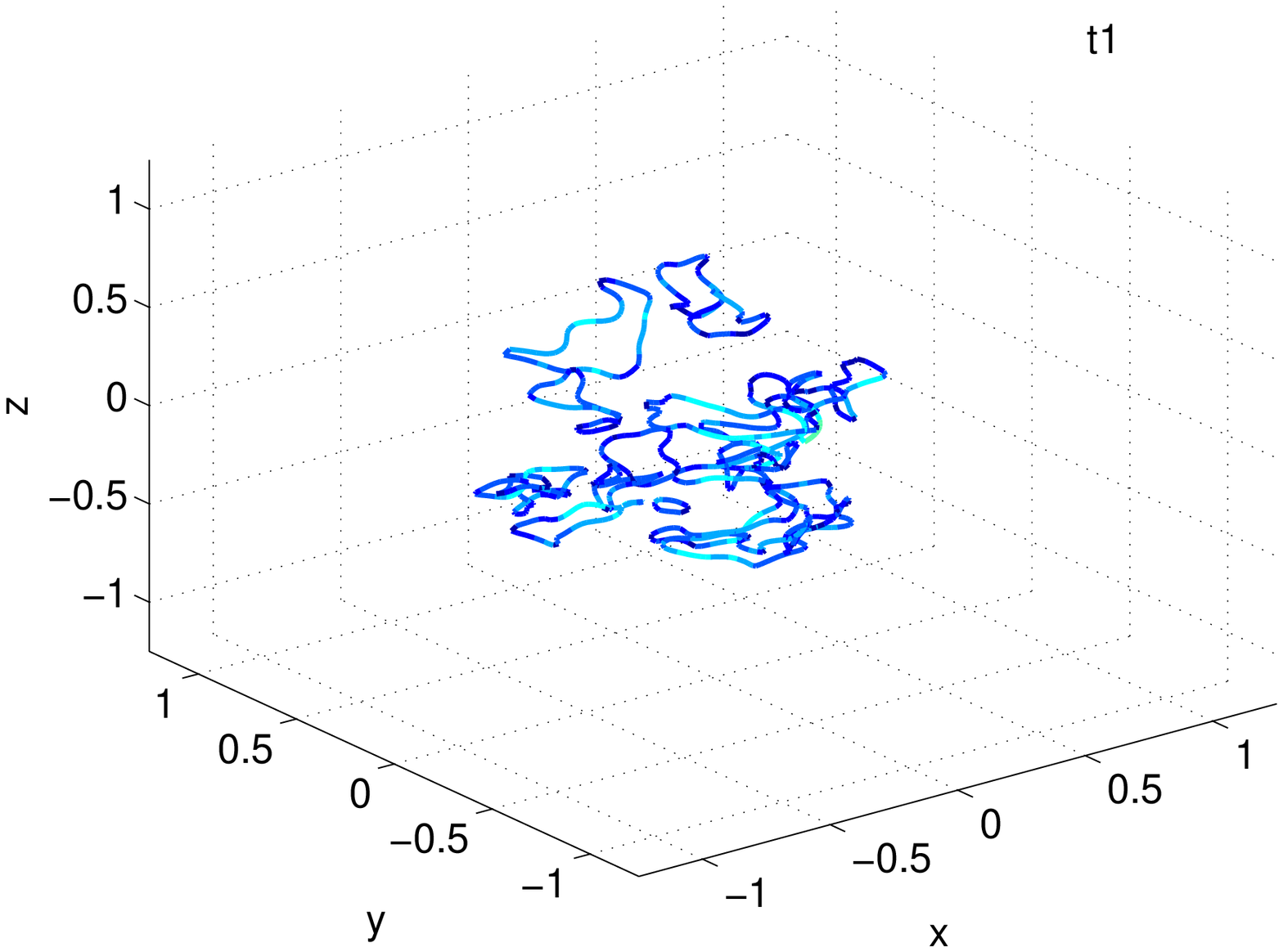}
\quad
    \includegraphics[width=0.4\textwidth]{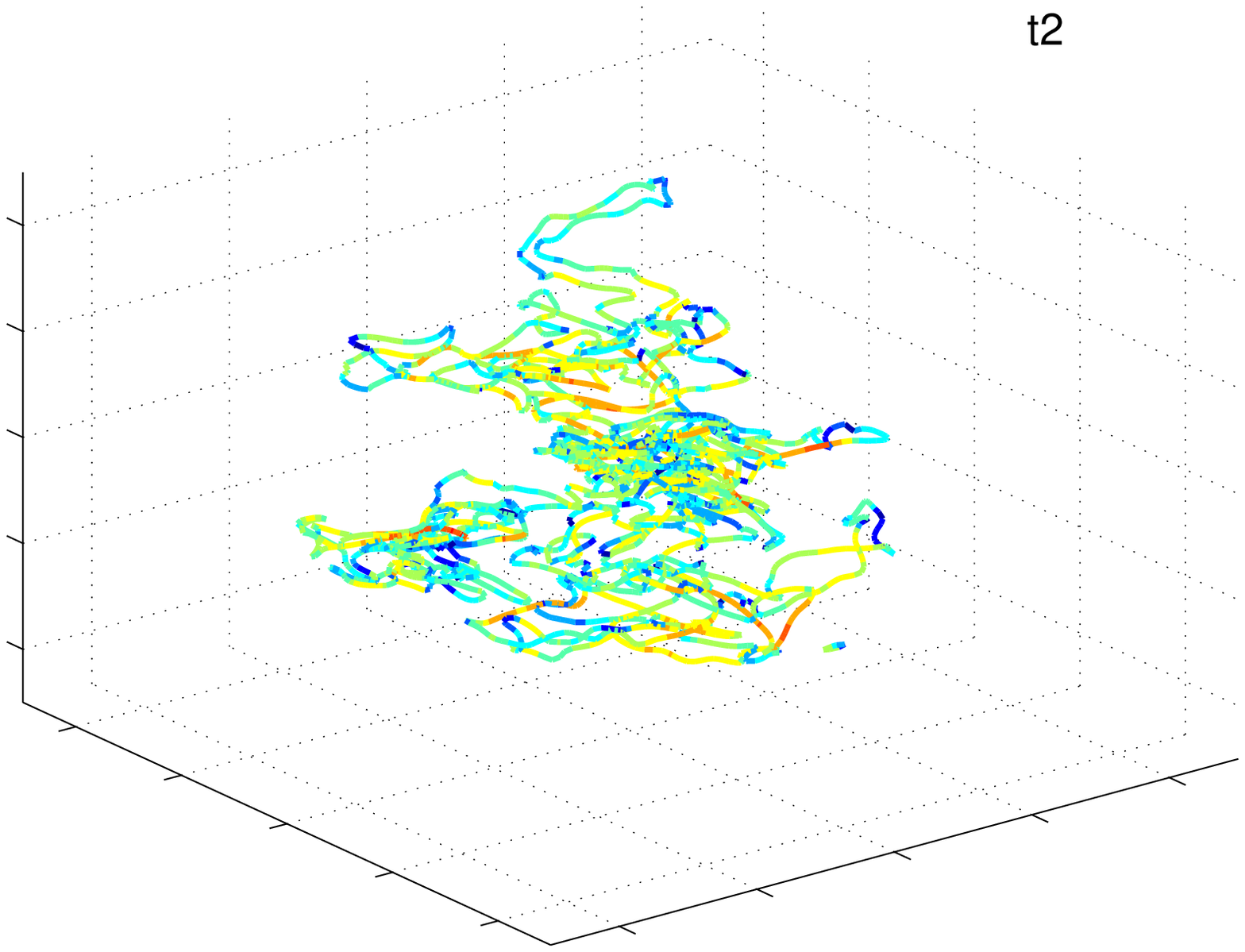}\\
    \includegraphics[width=0.4\textwidth]{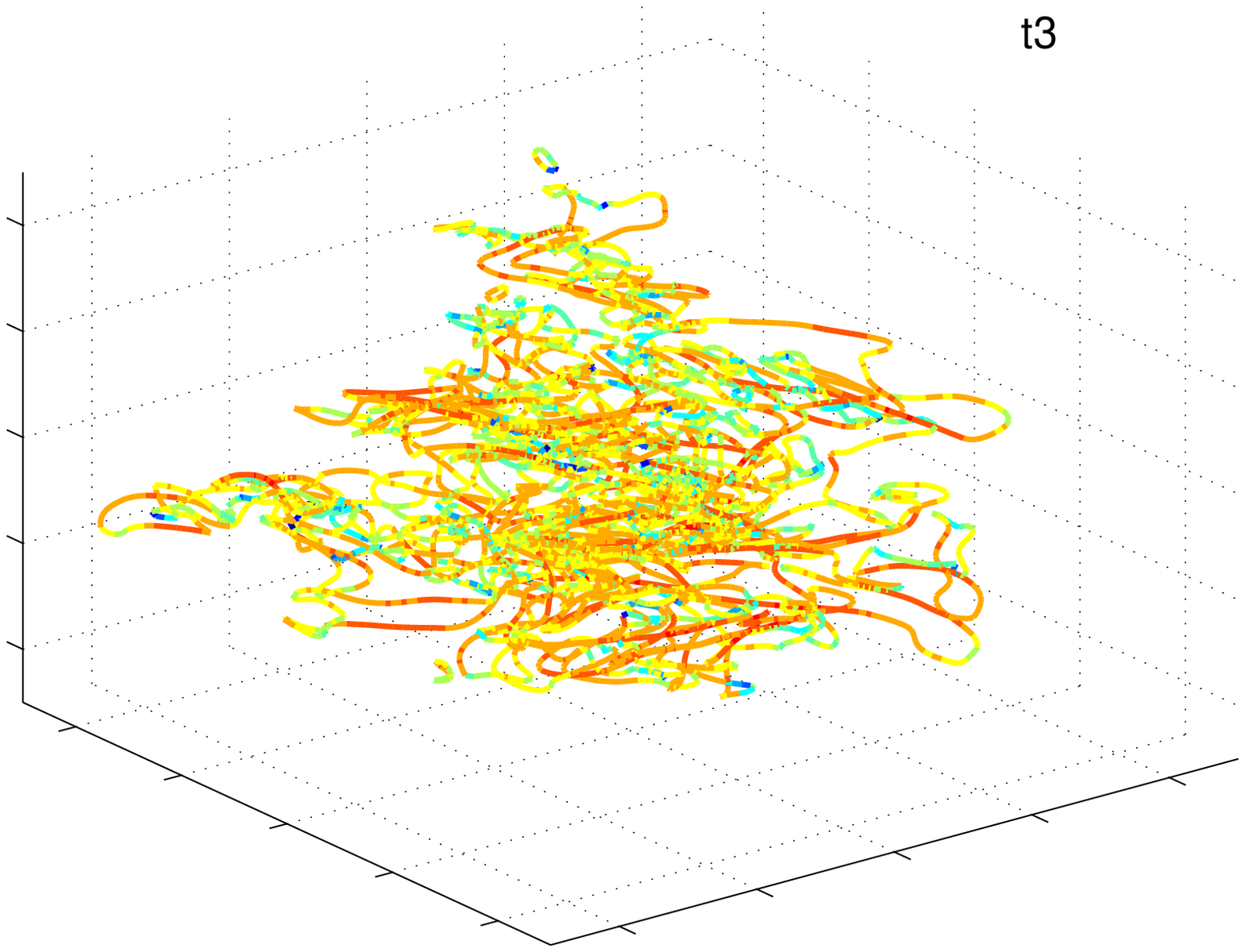}
\quad
    \includegraphics[width=0.4\textwidth]{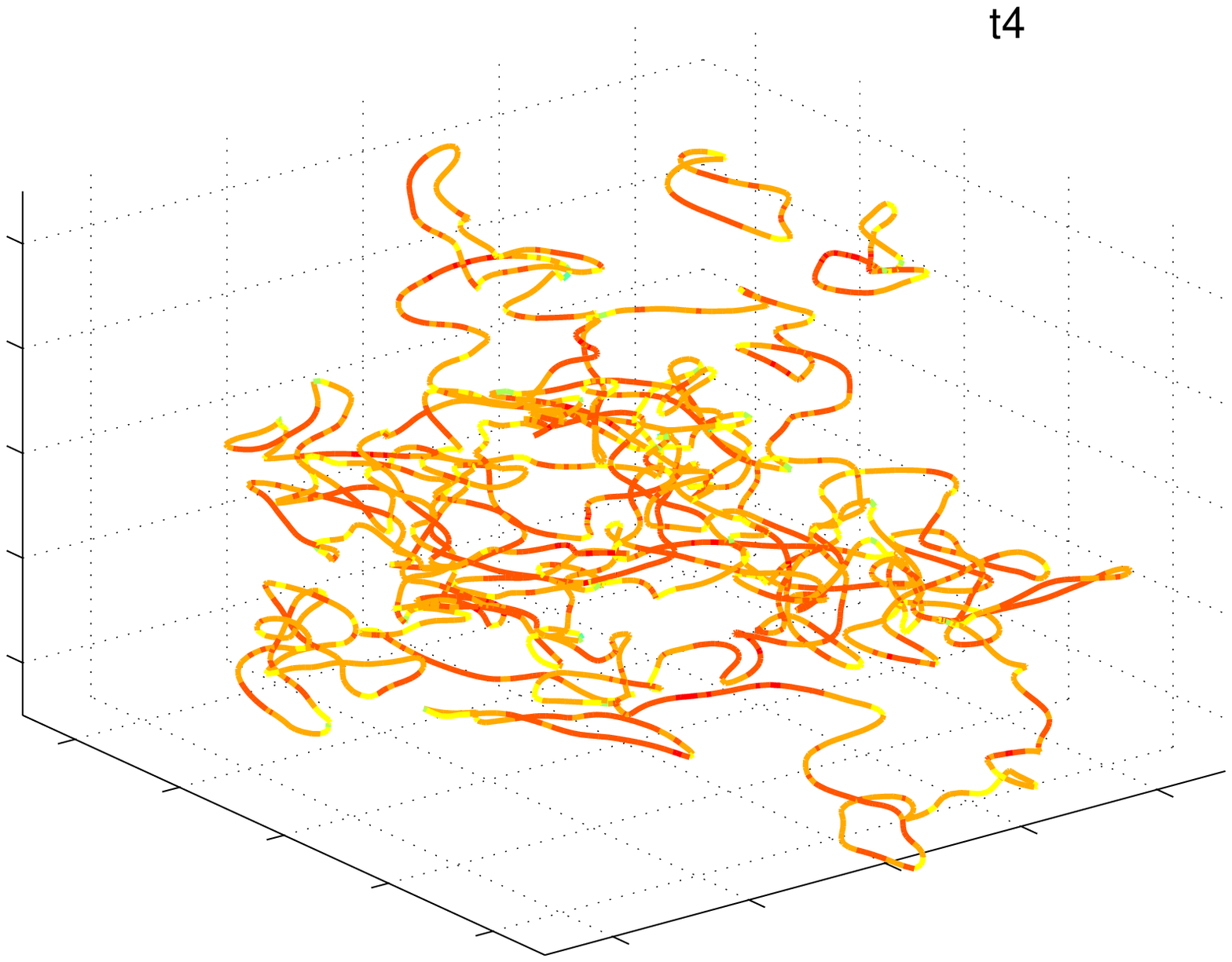}\\
    \includegraphics[width=0.4\textwidth]{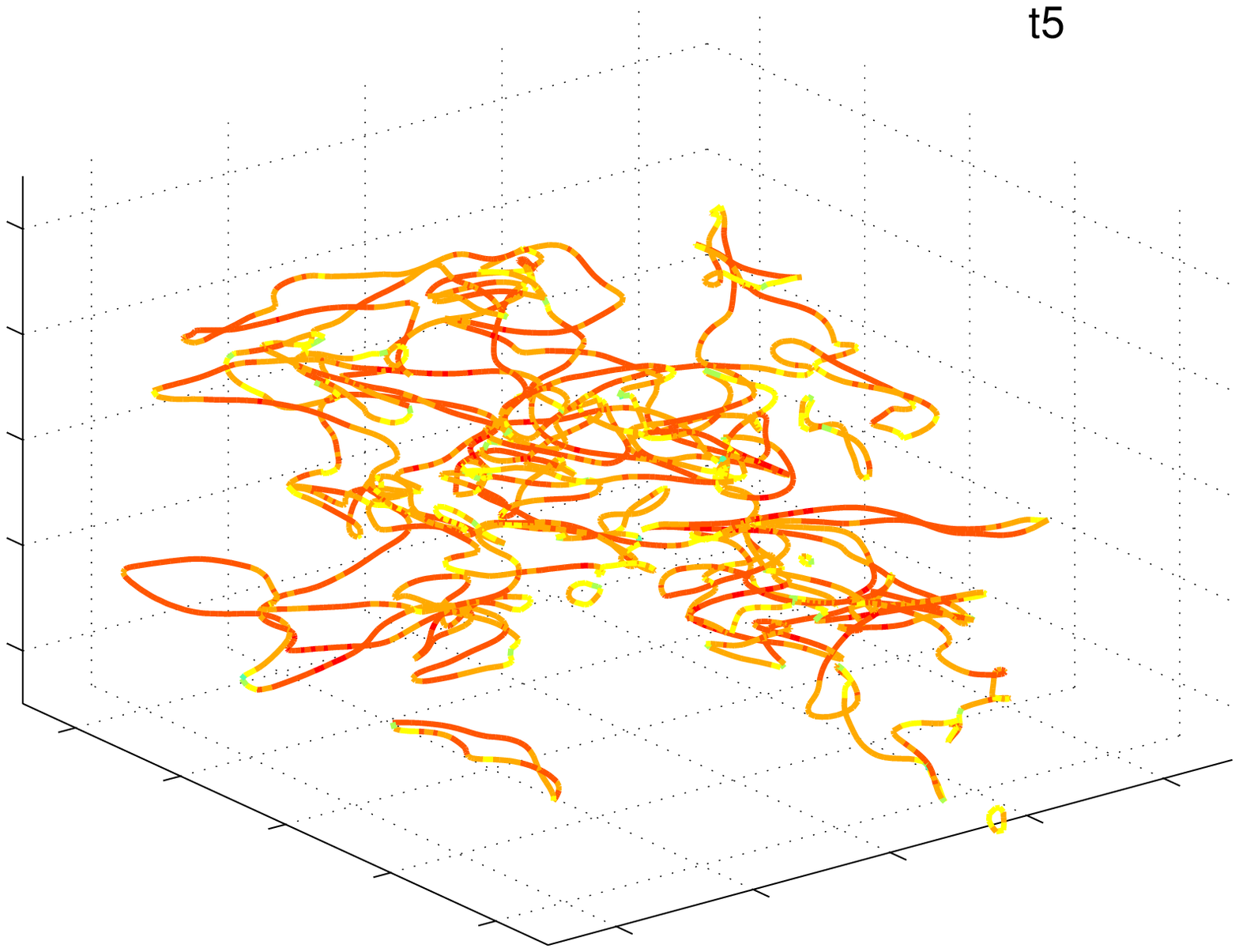}
\quad
    \includegraphics[width=0.457\textwidth]{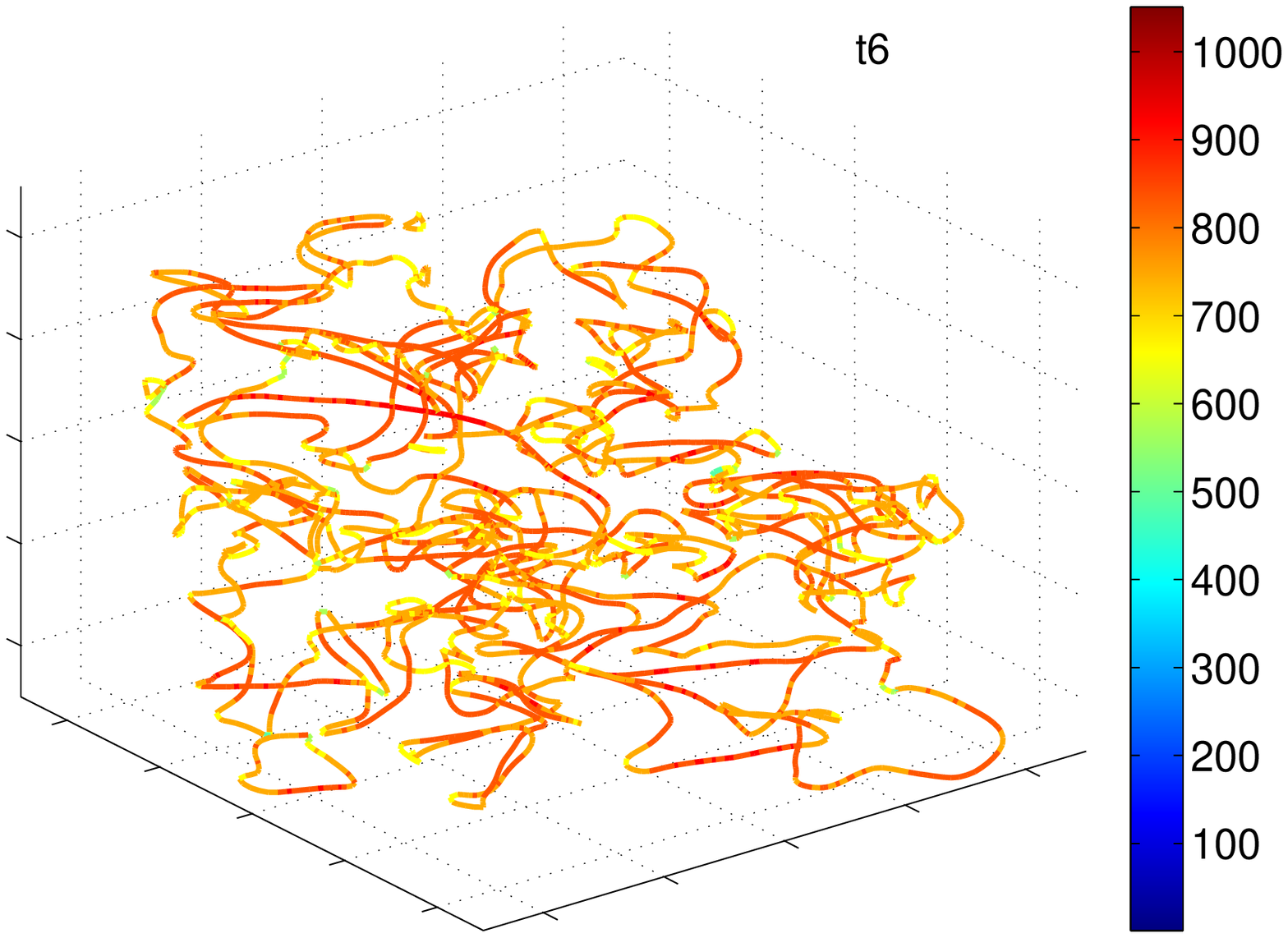}
    \caption{\label{KS_sat_snap}(Colour online) Snapshots of the magnetic
loops evolving in the KS flow (\protect\ref{uF}) in both the kinematic stage
$t\lesssim4$ and in the saturated state $t\gtrsim4$, taken from the same run as
Fig.~\protect\ref{saturated_brms}. Magnetic field strength in the flux tubes
is colour coded, with the colour scheme shown in the bottom right corner. Note the scale of the box remains the same (dimensions shown at $t=0.6$) for all snapshots.
Notice the high density of the snapshot when $t=3.9$ and the field is at a maximum.
This corresponds to an overshoot in $B_\mathrm{rms}$ visible in Fig.~\protect\ref{saturated_brms} as the dynamo saturates.}
\end{center}
\end{figure*}

Snapshots of the magnetic ropes shown in Fig.~\ref{KS_sat_snap} illustrate the
spatial evolution of the magnetised region. Unlike virtually all earlier
simulations of the fluctuation dynamo, most often performed in periodic boxes
with volume-filling initial conditions, the initial magnetic field in our
simulations is localised in space, as shown in the upper left panel of
Fig.~\ref{KS_sat_snap} (which refers to an early stage of evolution). In
the kinematic stage, $t\lesssim4$, magnetic field growth is accompanied by the
spread of the magnetised region clearly visible in the first three snapshots.
Consistently with the action of magnetic diffusion
$\eta_\mathrm{t}\propto l_0u_0$, the size
of the region occupied by magnetic ropes grows at $t^{1/2}$
However, the spread is halted in the nonlinear, saturated stage
represented in Fig.~\ref{KS_sat_snap} by snapshots at $t=25,\ 50$ and $85$,
which suggests that the turbulent magnetic diffusivity is suppressed in the
saturated state.
This appears to be a result of the suppression, by magnetic tension, of
random stretching of magnetic field at the location of a flux rope.
In other words, the saturation of the dynamo action is
achieved via the suppression of the effective magnetic Reynolds number,
$\Rmt=l_0u_0/\tilde{\eta}$, where $\tilde{\eta}$ is the effective
microscopic magnetic diffusivity.
This idea is fully consistent with the arguments of \citet{S99}
who considered a similar nonlinearity in the Kazantsev model of the fluctuation
dynamo \citep[see also][]{kim99,kim00,S03}
and also of \citet{Schek02,Schek04} who suggested a simple model of the effect
of Lorentz force.
A general feature of these models is that the small-scale dynamo saturates
because of a `renormalization' of the coefficients governing
its evolution, and the corresponding decrease in the
effective magnetic Reynolds number. This can be the result of
enhanced nonlinear diffusion \citep{S99},
increased diffusion together with additional hyperdiffusion \citep{S03},
or reduced stretching \citep{kim99,kim00,Schek02,Schek04}.
Our model is consistent with  the saturation
of the dynamo action via the suppression of the magnetic Reynolds number,
now arising from a reduction of localised random stretching, or turbulent
magnetic diffusivity.
\begin{figure}
  \begin{center}
    \psfrag{x}{$t/t_0$}
    \psfrag{y}{$\Delta_B$}
    \includegraphics[width=0.4\textwidth]{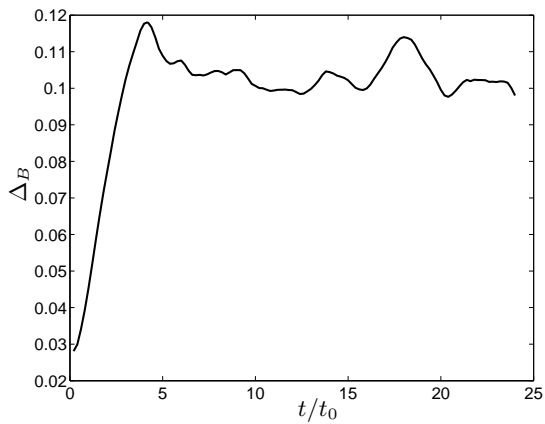}
    \caption{\label{dust}$\Delta_B$, the average separation of trace
particles arranged along magnetic loops in the KS flow (\ref{uF}), grows
rapidly in the kinematic regime of the dynamo, and then remains roughly
constant as soon as the dynamo action saturates at $t\gtrsim4$ when the
magnetic tension force becomes significant.}
\end{center}
\end{figure}

To clarify further the mechanism of dynamo saturation,  we monitored the
average separation of trace particles in the flow. At the start of a
simulation each trace particle located on a magnetic loop is assigned a
neighbour, for convenience the particle next to it. As the
simulation proceeds, the particles are advected by the flow, and new
particles may be introduced between them, but we continue to monitor the
separation between the original pair of particles. The particle separation
averaged over all the original particle pairs, $\Delta_B$, is shown in
Fig.~\ref{dust}. Indeed, the separation of the particles
stops growing as soon as the dynamo enters the nonlinear
stage. We stress that, at late times, not all pairs of trace particles
belong to the same magnetic loop because of multiple reconnections that often
split a magnetic loop into smaller ones. Thus, the fact that $\Delta_B$
ceases to grow implies that not only stretching is suppressed within a single
loop, but also that the magnetic loops stop spreading in space.

\begin{figure}
  \begin{center}
    \psfrag{t/t0}{ $t/t_0$}
    \psfrag{Delta}{ $\Delta_u$}
\includegraphics[height=0.4\textwidth,angle=-90]{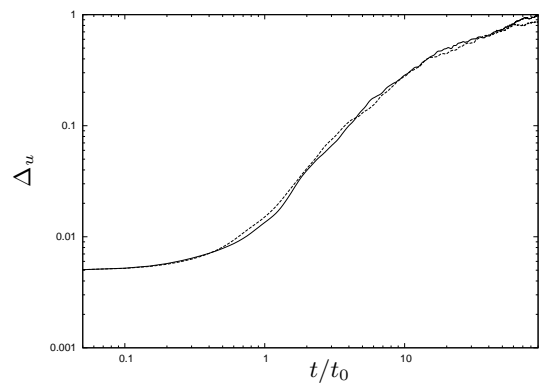}
    \caption{\label{sep}$\Delta_u$, the average separation of 1000 test
particles in snapshots of the KS flow (\ref{uF}), unaffected (solid) and
affected (dashed) by magnetic field.}
\end{center}
\end{figure}

We performed another experiment, where the location of the test
particles used to compute the dispersive properties of the flow was not
restricted to the magnetic loops. To reduce technical problems, we considered
two time-independent flows obtained as the snapshots of the original KS
velocity field and of its form affected by magnetic tension force at a certain
moment in the saturated dynamo state. The evolution of the particle separation
(averaged over 500 pairs) is shown in Fig.~\ref{sep}, where one can
distinguish the initial exponential growth of the separation,
    followed by the Richardson regime $\Delta_u\propto t^{3/2}$ when
    $\Delta_u \lesssim l_0$, and then by the incoherent, diffusive dispersion
$\Delta_u \propto t^{1/2}$ at larger separations. However,
the separation of the particles is insensitive to the effects of
the Lorentz force: the two curves in Fig.~\ref{sep} hardly differ from each
other. Thus, the flow has not been strongly affected by the Lorentz force, except
for the close vicinity of the flux tubes.

Importantly, the mean particle separation $\Delta_u$ is a measure of
\textit{kinetic\/} turbulent diffusivity, as opposed to the magnetic one. The
former involves the mean square of the {\it total\/} velocity
$\nu_\mathrm{t}\propto\langle \tau u^2\rangle$, whereas the turbulent magnetic
diffusivity is only sensitive to the velocity field components orthogonal to
the magnetic field, $\nu_\mathrm{t}\propto\langle \tau u_\perp^2\rangle$.
Incidentally, these results imply that the turbulent magnetic Prandtl number
is different from unity: in isotropic flow and magnetic field,
$\Pr_\mathrm{m}=\nu_\mathrm{t}/\eta_\mathrm{t}\simeq 3/2$; the difference is
small but perhaps significant in some applications. We cannot exclude that this feature
is an artifact of our model where a localised modification of the velocity
field by magnetic tension does not spread into a broader region as it would do
due to kinematic viscosity.

\begin{figure*}
  \begin{center}
    \psfrag{x}{x}
    \psfrag{y}{y}
    \psfrag{t1}{t=0.0}
    \psfrag{t2}{t=1.5}
    \psfrag{t3}{t=5.0}
    \psfrag{t4}{t=10.0}
    \psfrag{t5}{t=13.8}
    \psfrag{t6}{t=14.8}
    \includegraphics[width=0.4\textwidth]{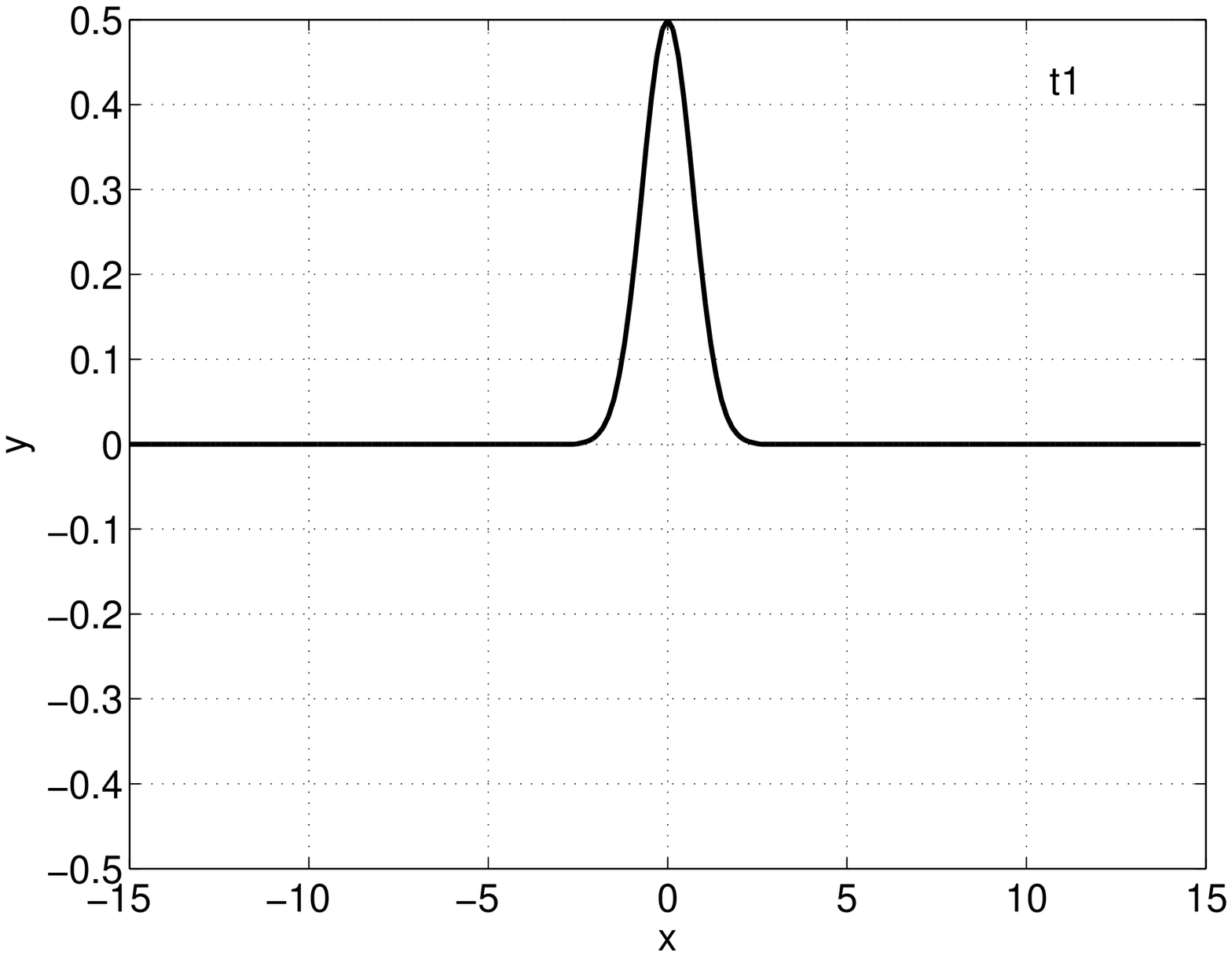}
    \includegraphics[width=0.4\textwidth]{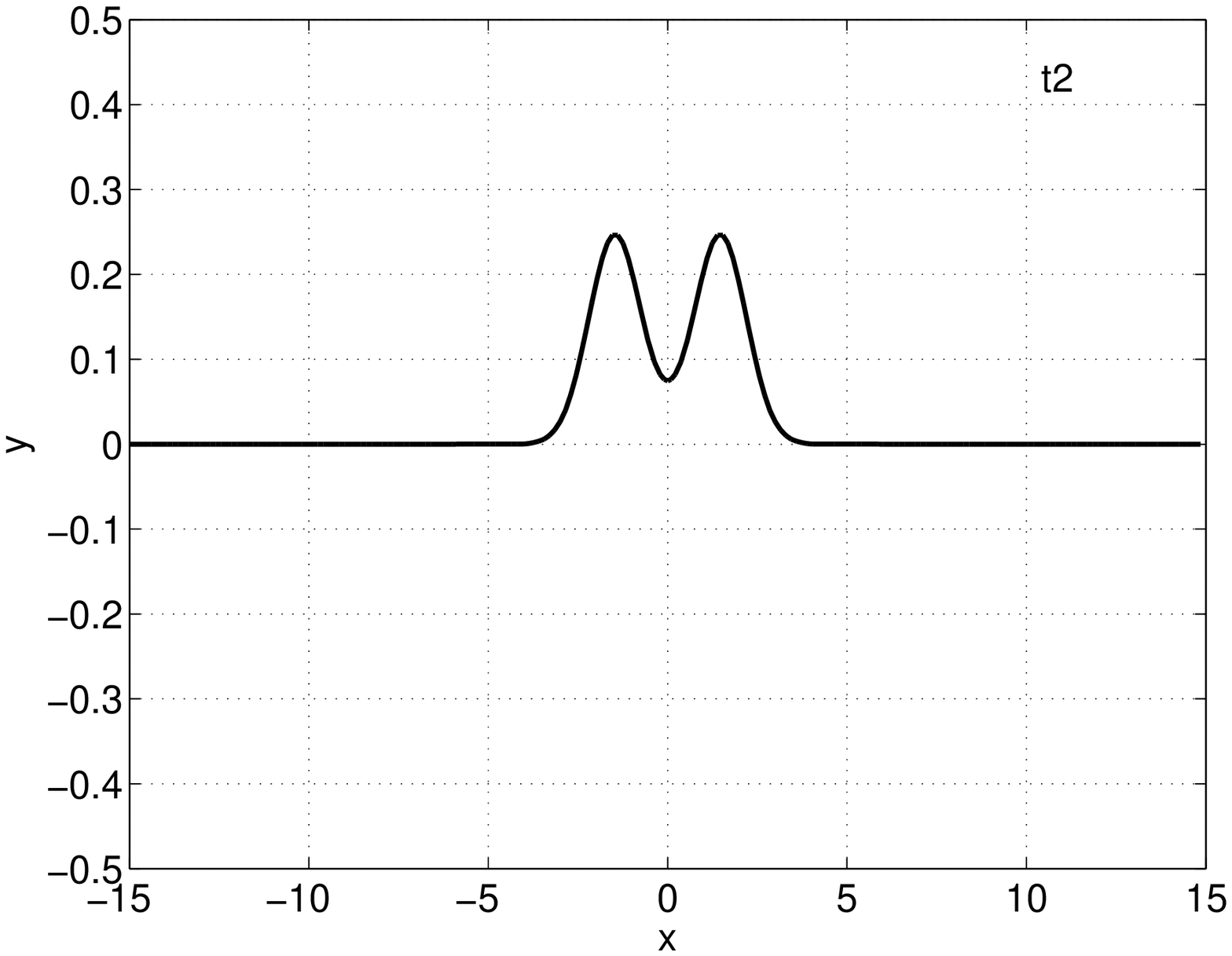}
    \includegraphics[width=0.4\textwidth]{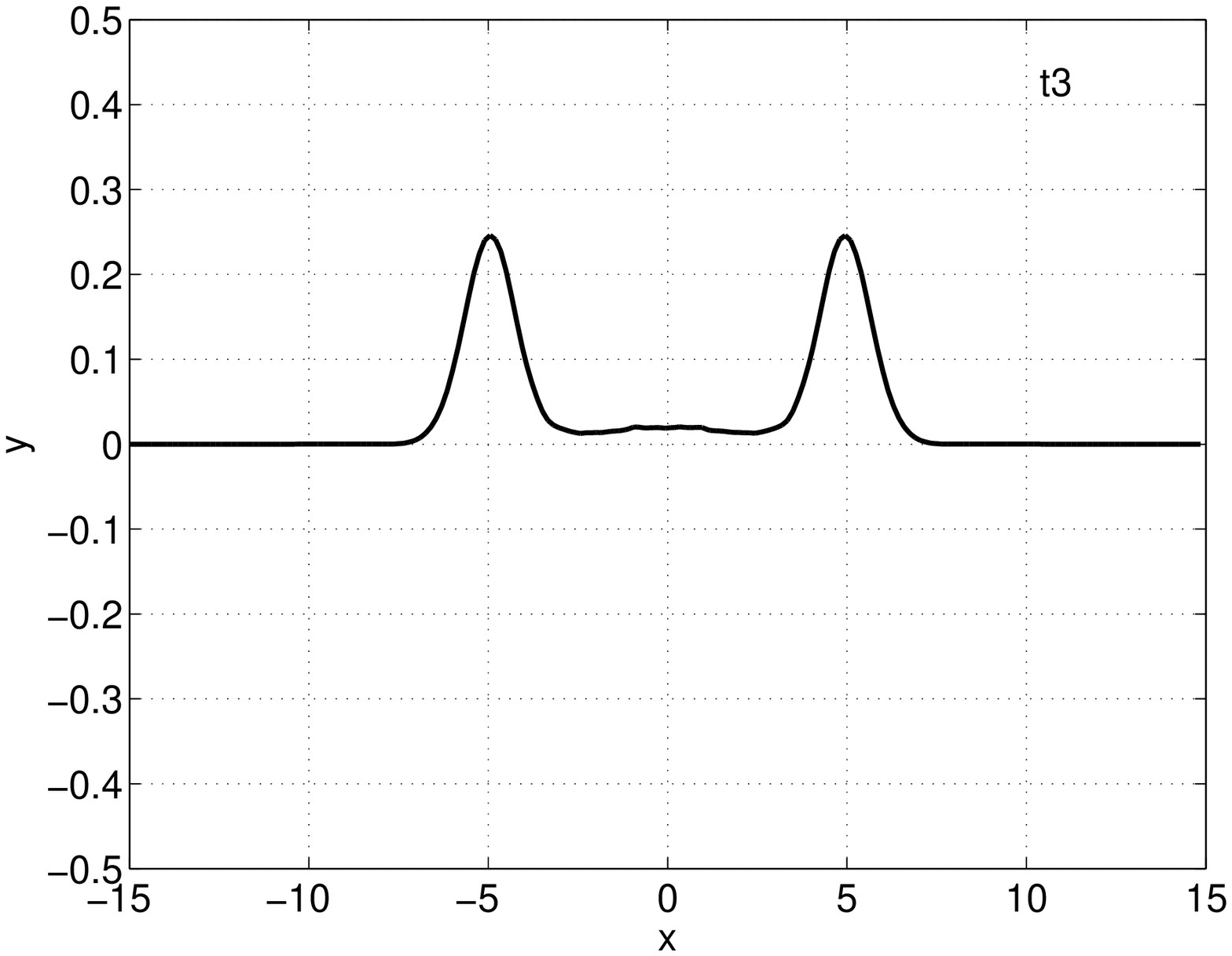}
    \includegraphics[width=0.4\textwidth]{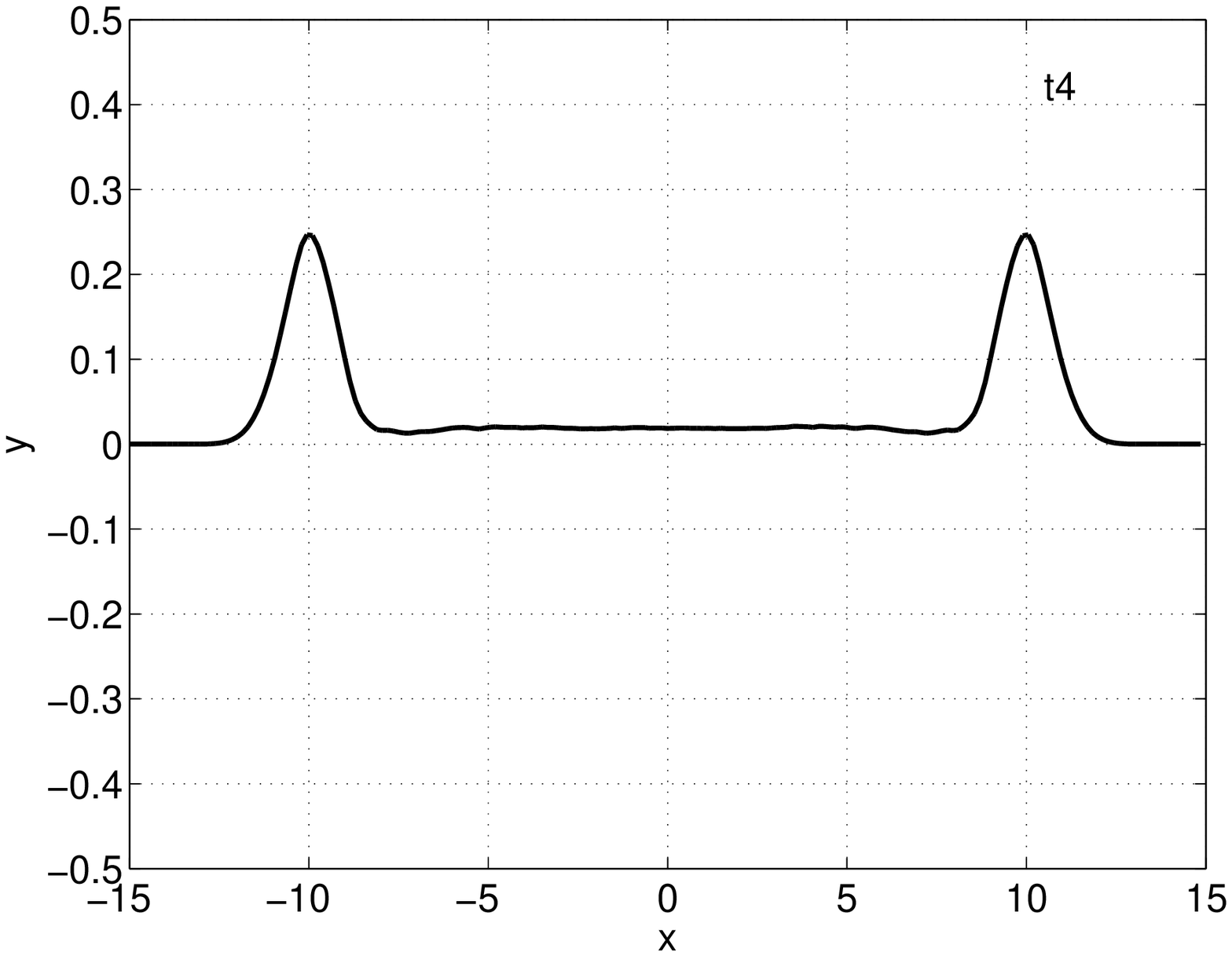}
    \includegraphics[width=0.4\textwidth]{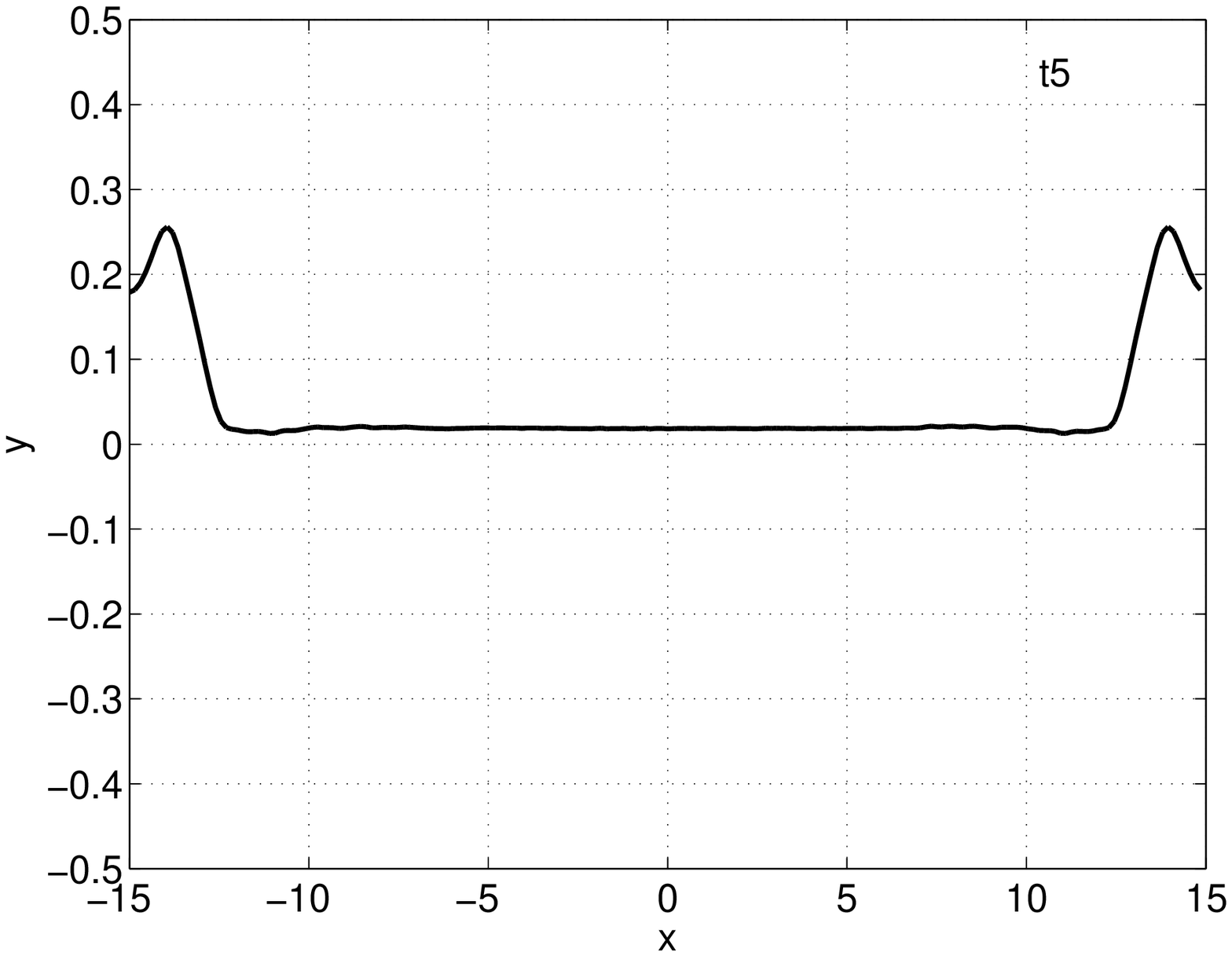}
    \includegraphics[width=0.4\textwidth]{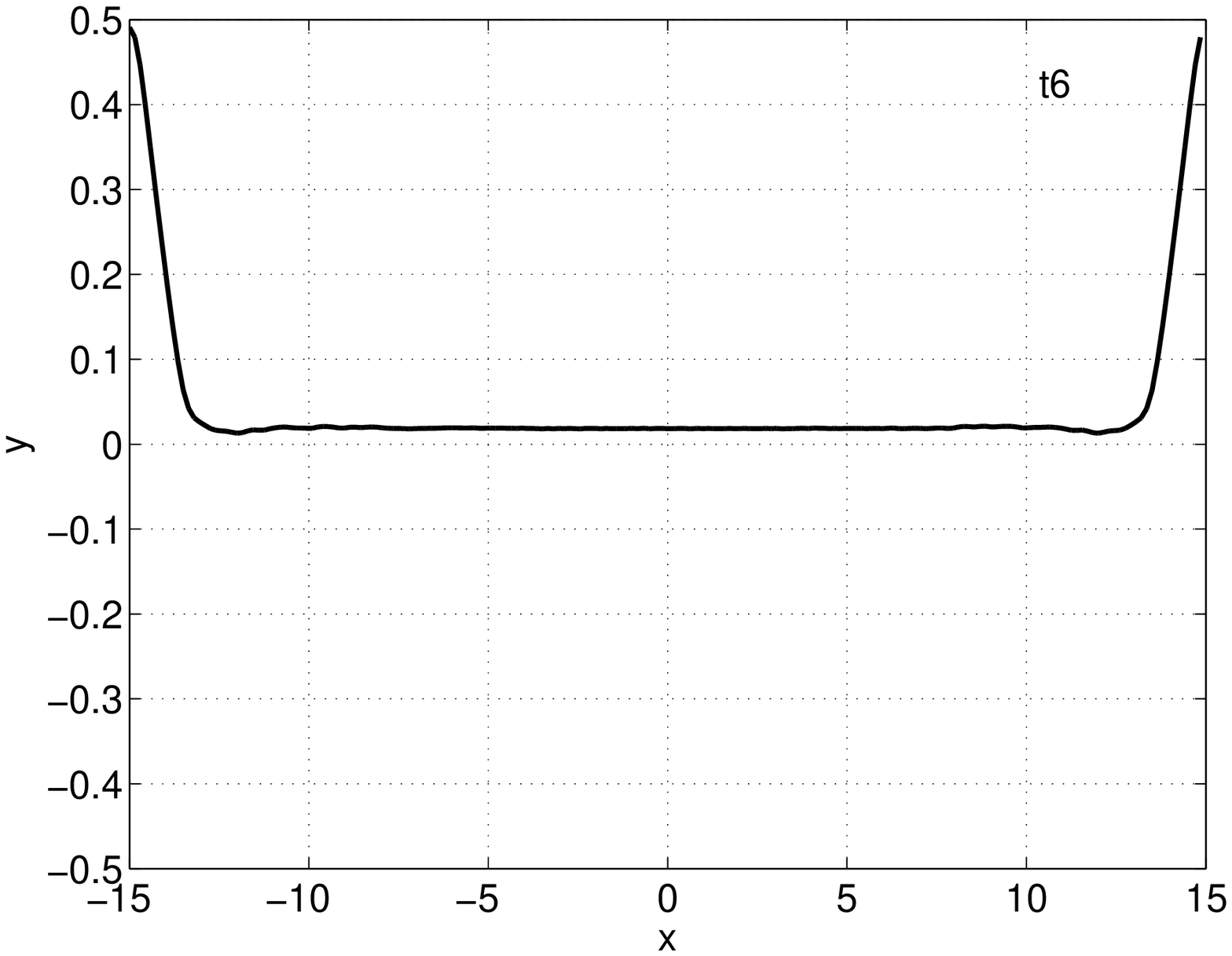}
    \caption{\label{alfven_snap}Snapshots of the Alfv\'en wave simulations. A
single flux tube with field strength $|\mathbf{B}|=1$ is perturbed at the
midpoint. As the simulation progresses two wavefronts form which move apart
with speed equal to the field strength. As the fronts reach the edge of the
periodic box they interact at the boundary.}
\end{center}
\end{figure*}

\subsection{Alfv\'en waves}
To demonstrate the flexibility of the flux model suggested here, we briefly
present simulations of Alfv\'en waves propagating along an imposed magnetic
field. For this purpose we solve Eqs.~(\ref{fullmodelnoks}) and
(\ref{bvolution}) with a single magnetic flux tube which, initially, has
constant magnetic field. The tube is perturbed as shown in the upper left
panel of Fig.~\ref{alfven_snap}.  The
simulation proceeds in a box periodic in the $x$-direction, and the
interaction with the ghost wave from the next periodicity cell
is seen in the final snapshot at $t=14.8$. We confirmed that the phase speed
of the wave is indeed proportional to the strength of the magnetic field.
Since our model admits nonlinear interactions of Alfv\'en waves, it can be
used to study spectral energy cascades and other features of the Alfv\'en wave
turbulence.

\section{Curvature of magnetic lines}
\citet{Sheck:2002} discuss the geometry of magnetic lines in the kinematic
fluctuation dynamo driven by a single-scale, $\delta$-correlated in time
random flow with high magnetic Prandtl number. They argue that
magnetic field strength and magnetic line curvature should be anticorrelated
and derive a power-law probability distribution function of the
magnetic line curvature. These results are used to support the picture of
folded magnetic lines as a representation of magnetic field produced by the
fluctuation dynamo.

The anticorrelation between the curvature of magnetic lines and the
strength of the magnetic field is intuitively appealing since magnetic field
strength grows due to a random stretching of magnetic lines which
is necessarily accompanied by a reduction in their local curvature. However,
the stretching is not the only component of the fluctuation dynamo mechanism.
In the framework of the stretch-twist-fold dynamo concept, stretching must
be followed by the folding of magnetic lines to ensure an exponential growth
of magnetic field -- and the folding will tend to increase the local magnetic
line curvature. Therefore, an anticorrrelation between magnetic curvature and
strength may be expected for a decaying magnetic field rather than for
magnetic fields growing due to the dynamo action. In this section we explore
directly the relation between the magnetic line curvature and strength using
the reconnecting flux rope dynamo model.

The curvature of the flux ropes can be calculated as \citep{gray:1996}
\begin{equation}\label{curvature_eqn}
\kappa=\frac{|\mathbf{r}' \times \mathbf{r}''|}{|\mathbf{r}'|^3},
\end{equation}
where $\mathbf{r}(s)$ is a parametrised space curve representing a magnetic
flux rope, with $s$ the distance measured along the rope, and dash denotes
derivative with respect to $s$, with the first derivative calculated using
Eqs.~(\ref{first_deriv}) and the second derivative, from
\begin{equation}\label{second_deriv}
  \mathbf{r}_i''=\frac{2\mathbf{r}_{i+1}}{\ell_i(\ell_i+\ell_{i-1})}
        -\frac{2\mathbf{r}_i}{\ell_i\ell_{i-1}}
        +\frac{2\mathbf{r}_{i-1}}{\ell_{i-1}(\ell_i+\ell_{i-1})}+O(\ell^2),
\end{equation}
where notation is defined in Fig.~\ref{derive_scheme}. For
$\ell_i=\ell_{i-1}=h$, this form reduces to a standard finite difference
scheme.

\begin{figure}
  \begin{center}
    \psfrag{x}[c]{$\log |\mathbf{B}|/B_\textrm{max}$}
    \psfrag{y}[c]{$\kappa/\kappa_\textrm{max}$}
    \vspace{5mm}
    \includegraphics[width=0.4\textwidth]{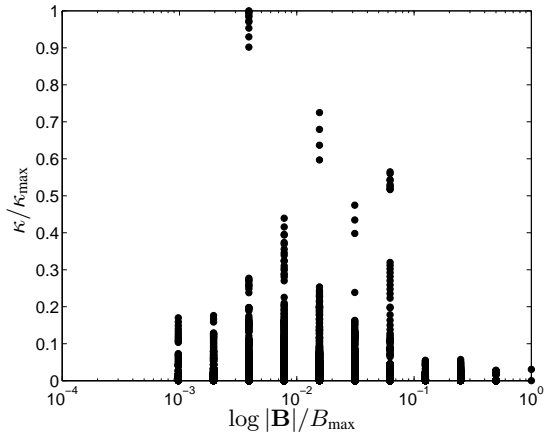}
    \caption{\label{curv1} The scatter plot of curvature, $\kappa$
versus $|\mathbf{B}|$ calculated at the end of a simulation, with both variables
normalised to the respective maximum values.}
\end{center}
\end{figure}

Figure~\ref{curv1} is the scatter plot of magnetic field strength at a
particular position versus the field line curvature at that point, computed at
the end of the simulation, illustrated in Fig.~\ref{KS_snap}, where the dynamo
is driven by the KS flow. Only the envelope of the curvature distribution
appears to be consistent with the anticorrelation, and even that only for relatively
strong fields, $B\gtrsim10^{-2}B_\mathrm{max}$. Thus, the range of the curvature
values is narrower at positions where the field is stronger, but for any field
strength this range includes very small curvature values.

Those parts of magnetic flux tubes
where magnetic field is weak have low curvature, especially those with $B\ll
B_0$ with $B_0$ the initial field strength
($B_0\approx4\times10^{-3}B_\mathrm{max}$ at the particular time of the
simulation).
In our model, the only way the field strength $|\mathbf{B}|$ can become smaller
than $B_0$ is through the shrinking of a flux tube caused by contracting flow.
In a perfectly conducting fluid, such a contraction can make the curvature
larger, e.g., when a wavy magnetic line is contracted along its wave vector.
However, the situation changes entirely in the presence of reconnections (or
any other magnetic dissipation mechanism): now, reconnections eventually
eliminate the bends of the magnetic line thus reducing the curvature of a
contracting magnetic line. Apparently, we see the evidence of this in
Fig.~\ref{curv1}. The group of points with nearly maximum curvature
at the top of the frame are probably those which will undergo reconnections of this
type very soon.
Finally we note that the reconnection length $d_0$ limits the maximum
value that $\kappa$ can take. 

\begin{figure}
  \begin{center}
    \psfrag{y}[c]{$\ln P(\kappa)$}
    \psfrag{x}[c]{$\ln \kappa$}
    \psfrag{u}[l]{$\kappa^{-13/7}$}
    \includegraphics[width=0.4\textwidth]{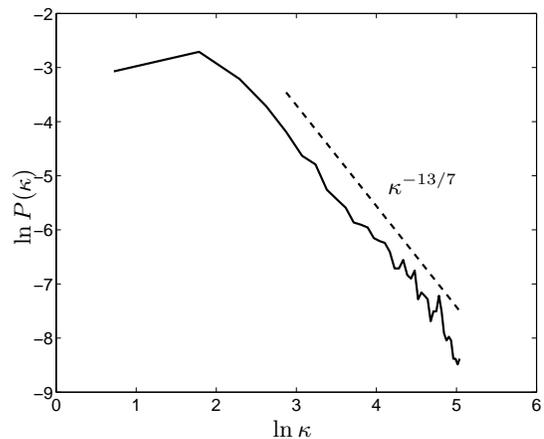}
    \caption{\label{curv_PDF} Probability density functions (PDF) of
curvature, $\kappa$ at a late stage of magnetic field evolution in the KS flow
    illustrated in Fig.~\ref{KS_snap}. Dashed line is for the power-law
distribution $P(\kappa)\propto \kappa^{-13/7}$ obtained by \citet{Sheck:2002}.}
\end{center}
\end{figure}

\citet{Sheck:2002} showed, both analytically and numerically,
that the probability density function $P$ of the curvature of
field lines has a power-law form in the limit of large $\kappa$. In
particular, they obtain $P(\kappa)\propto\kappa^{-13/7}$ for a
three-dimensional, incompressible flow. Figure \ref{curv_PDF} shows the PDF
of curvature from our simulations which shows a very good agreement
with the analytical results of \citet{Sheck:2002b}.

\begin{figure}
  \begin{center}
    \psfrag{x}{$t/t_0$}
    \psfrag{y1}{$\alpha$}
    \psfrag{y2}[c]{$\log B_\mathbf{rms}$}
    \includegraphics[width=0.4\textwidth]{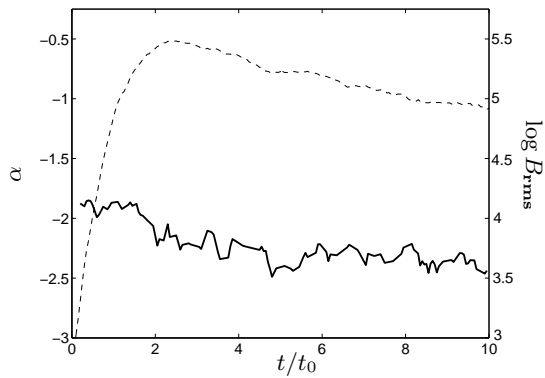}
    \caption{\label{curv_PDF_Brms} The slope of the PDF of magnetic line curvature
and r.~m.~s.~ magnetic field in the kinematic and saturated states of the dynamo. The solid 
line shows $\alpha$, where PDF$(\kappa) \sim
\kappa^{\alpha}$. In the kinematic regime $\alpha \approx -13/7$
as in Fig.~\protect{\ref{curv_PDF}}. As the dynamo saturates 
($\log B_\mathrm{rms}$ shown dashed) $\alpha$ decreases.}
  \end{center}
\end{figure}

\citet{Sheck:2002b} argue that, because the magnetic field strength is higher
where the magnetic line curvature is lower (due to the stretching by velocity
shear), magnetic tension force (which is quadratic in $\mathbf{B}$) is lower
at positions with weak field. This conclusion relies on
estimating the tension force as $|B\partial\mathbf{B}/\partial s| \simeq
\kappa B^2$, and noting that this quantity decreases with $B^2$, although $\kappa$ increases. 
This would imply that magnetic tension
is relatively unimportant in regions with
strong curvature. This leads these authors to a conclusion that magnetic
field is organised into folded structures which persist in the saturated
state. However, what matters is not the value of $B^2$ itself but rather its
gradient along the magnetic line, $\partial B^2/\partial s$. 
In a stretched magnetic line, the gradient is
reduced in regions with strong magnetic field (i.e., the straight segments of
the folded magnetic lines) and enhanced in regions of weaker field (in the
turns in the folded structures). Therefore, magnetic tension will drive the
turns closer to each other along each magnetic line destroying the folded
structures. Thus, the lack of any apparent domination of folded structures in
Fig.~\ref{KS_snap} is consistent with the curvature PDF shown in
Fig.~\ref{curv_PDF}. We show in Fig.~\ref{curv_PDF_Brms} the time variation of
the slope of the curvature PDF into the nonlinear regime (discussed in
Section~\ref{NLM}): the PDF becomes steeper , so that high curvature occurs less often
in the nonlinear state. This can be attributed to magnetic tension which tends to reduce magnetic line curvature.
\section{Statistics of magnetic energy release}
Solar corona is one of the astrophysical environments where magnetic
reconnections are believed to play important role, particularly in heating the
plasma to the high temperatures observed \citep{priest:2000}. The
reconnections are assumed to be driven by the motion of the footpoints of
magnetic flux tubes anchored in the photosphere and extending into the corona
\citep[][and references the\-re\-in]{PLT03}. Reconnection events
that release large amounts of magnetic energy are observed as solar flares. A
remarkable feature of the coronal heating mechanism is that the frequency
distribution of the flare energy has a power law form in a very broad energy
range (eight orders of magnitude) \citep[see an excellent review of][]{SOC}
\begin{equation}\label{pdf}
P(\Delta M)\propto (\Delta M)^{s}\;.
\end{equation}
If $s<-2$, most of the magnetic
energy released into the corona is due to weak flares. This attractive option
suggested by \citet{P83} is known as the nanoflare model of the coronal
heating. This idea is most often
explored in the context of
self-organised criticality models based on cellular automata, which are known
to demonstrate the required power-law statistical distributions. Notably, the
continuous analogies of these models involve
the
 hyperdiffusion operator
\citep{SOC}. A widely recognised difficulty of this approach
is the elusive connection with the physical picture and even unclear physical
interpretation of the variables. Alternative models
\citep[e.g.,][]{Hughes:2003}, where reconnection evens are modelled directly,
also reproduce the power-law statistics, but still remain rather idealised
regarding the behaviour of magnetic flux tubes.

Our model is quite different from the Solar corona
settings, where the reconnections are driven by the motion of the flux rope
footpoints, the plasma is believed to be magnetically dominated, and
\textit{in situ\/} dynamo action is improbable. Nevertheless, in this
section we consider the statistics of the energy release in our model of the
flux rope dynamo. As we show here, our reconnection dynamo model naturally
develops a power-law distribution (\ref{pdf}) with $s\simeq-3$, which appears
to be independent of the form of the velocity field. Our model can readily be
adapted to the Solar corona conditions, and despite the differences of our
model from the Solar corona models, we feel that this feature of the model can
be relevant in this context.

In the case of the induction equation, the magnetic energy dissipation rate
can be defined as
\begin{equation}
\gamma_\mathrm{i}=\frac{1}{M}\frac{dM}{dt}
=\eta \frac{\int_V \mathbf{B} \cdot \nabla^2 \mathbf{B}\, dV}{\int_V \mathbf{B}^2\, dV}\;,
\end{equation}
where $M$ is the total magnetic energy.
A similar quantity can be obtained for the flux rope dynamo by summing
the contributions of all reconnection events to the magnetic energy release:
\begin{equation}
\gamma_\mathrm{r}=\frac{1}{M}\frac{dM}{dt}=
{\frac{1}{8\pi M\tau}\displaystyle \sum_{i=1}^{N_\tau}B_i^2S_iL_i}\;,
\end{equation}
where $\tau$ is a suitable time interval during which $N_\tau$ reconnections
occur (we take $\tau$ to be equal to ten time steps; individual reconnection
events occur in a single time step), and $B_i$, $S_i$ and $L_i$ are the
magnetic field strength, the cross-sectional area and length of the
reconnected (and thus removed) flux tube segment associated with a trace
particle number $i$. From our assumption of frozen flux $B_i
S_i=\psi=\mbox{const}$, the total magnetic energy $M$ is,
\begin{equation}
M=\displaystyle \sum_{i=1}^{N_\mathrm{tot}}\frac{B_i^2}{8\pi}S_iL_i
= \frac{\psi}{8\pi}\sum_{i=1}^{N_\mathrm{tot}}B_iL_i\;,
\end{equation}
where $N_\mathrm{tot}$ is the total number of
trace particles
in all flux tubes.
Thus,
\begin{equation}
\gamma_\mathrm{r}=\frac{1}{\tau}
        \frac{\sum_{i=1}^{N_\tau}B_iL_i}{\sum_{i=1}^{N_\mathrm{tot}}B_iL_i}\;.
\end{equation}

\begin{figure}
  \begin{center}
    \psfrag{E}[c][position=1mm]{\vspace{1mm} $\gamma_\mathrm{i} l_0/u_0,
                                        \ \ \gamma_\mathrm{r} l_0/u_0$}
    \psfrag{t}[c][position=1mm]{\vspace{1mm} $t u_0/l_0$}
    \includegraphics[width=0.4\textwidth]{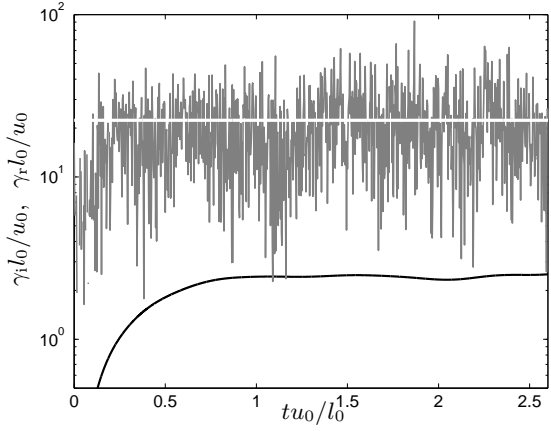}
    \caption{\label{combind_energy} Magnetic energy release rates from two
kinematic dynamo models with the KS flow and similar growth rates of magnetic
field: as obtained from the induction equation (black) and the reconnecting
flux rope model (grey). The former has a mean value of 2.4 (here
$\Rm=1200$) once the eigensolution has developed. The latter (with
$\tilde{R}_\textrm{m}=174$) has a mean value of 23 (shown with thick white
horizontal line).}
\end{center}
\end{figure}
\begin{figure}
  \begin{center}
    \psfrag{E}[c][position=1mm]{\vspace{1mm} $\gamma_\mathrm{i} l_0/u_0,
                                        \ \ \gamma_\mathrm{r} l_0/u_0$}
    \psfrag{t}[c][position=1mm]{\vspace{1mm} $t u_0/l_0$}
    \includegraphics[width=0.4\textwidth]{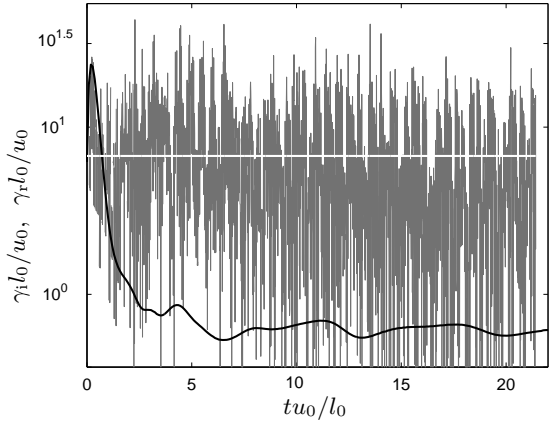}
    \caption{\label{combind_energy_ABC} As in
Fig.~\protect\ref{combind_energy}, but for the ABC flow with
$\gamma_\mathrm{i}l_0/u_0\approx 0.6$ for the eigensolution of
the induction equation at $\Rm=55$, and $\gamma_\mathrm{r}l_0/u_0\approx
6.7$ with $\tilde{R}_\mathrm{m}=24$ in the flux rope dynamo.}
\end{center}
\end{figure}

In Fig.~\ref{combind_energy}  we present the energy release rates in simulations
where the growth rate of the magnetic field is $\sigma=0.16$ in both
simulations (with the unit time $l_0/u_0$). The dashed line shows the energy
release rate from a simulation of induction equation with $\Rm=1200$, which
has the mean energy release rate $\gamma_\mathrm{i}\approx2.4$. The solid line
shows the corresponding results from the reconnection dynamo, with the mean
value plotted as a thick horizontal line. The mean value of the energy release
rate from the reconnecting flux rope dynamo is $\gamma_\mathrm{r}\approx23$,
an order of magnitude larger than that obtained from the induction equation.
We also note the strong fluctuations in the energy release rate from the
reconnection model, as opposed to the quiescent behaviour in the induction
equation. It is important that an order of magnitude difference in the 
energy release rates occurs in solutions with similar growth rates of magnetic field.
Since the reconnection based dynamo is more efficient than that based on magnetic diffusion
(see section \ref{DiffvsRecon}) kinetic energy density in the former being 10 times
smaller than in the latter.
With comparable kinetic energy densities, the difference between the energy release rates
can be even larger.

As shown in Fig.~\ref{combind_energy_ABC}, dynamos driven by the ABC flow
behave similarly. With $\Rm=55$, the induction equation gives an energy
release rate of about $\gamma_\mathrm{i}=0.6$. The corresponding flux rope
dynamo with the same growth rate ($\sigma=0.02$) has energy release rate
$\gamma_\mathrm{i}\approx6.7$, again ten times larger.

\begin{figure}
  \begin{center}
    \psfrag{z}[c]{$\log\zeta$}
    \psfrag{p}[l]{$\log P(\zeta)$}
    \includegraphics[width=0.4\textwidth]{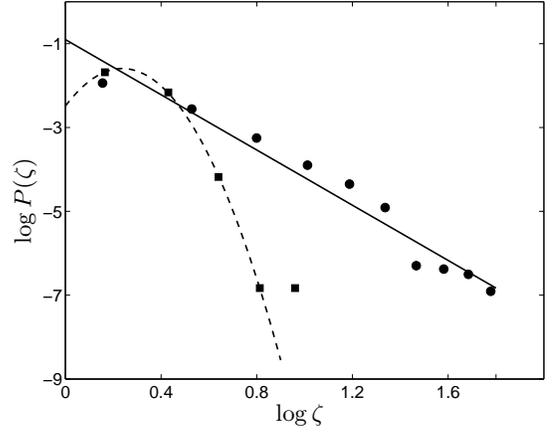}
    \caption{\label{pdf_KS} Probability density for the scaled magnetic energy
release, $\zeta=(\gamma-\overline{\gamma})/\sigma_\gamma$, from the time series
of Fig.~\protect\ref{combind_energy}, for the flux rope dynamo (circles) and
the diffusive dynamo with the same magnetic field growth rate and velocity field
of the same form (squares). A power-law fit
to the former and a Gaussian fit to the latter are shown solid and dashed,
respectively.}
  \end{center}
\end{figure}
\begin{figure}
  \begin{center}
    \psfrag{z}[c]{$\log \zeta$}
    \psfrag{p}[c]{$\log P(\zeta)$}
    \includegraphics[width=0.4\textwidth]{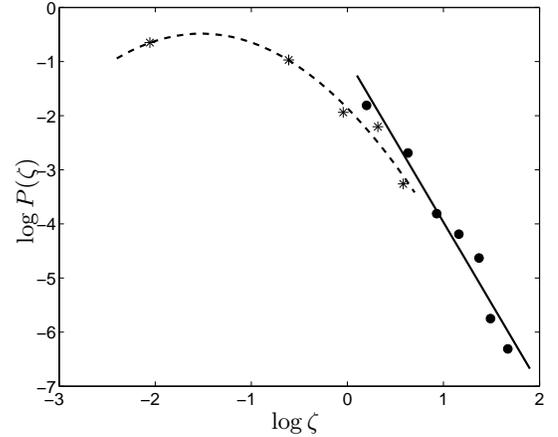}
    \caption{\label{pdf_ABC} As in Fig.~\protect\ref{pdf_KS}, but
from the time series of Fig.~\protect\ref{combind_energy_ABC}, a log-log plot
for the flux rope dynamo (circles) with solid line having the slope $-2.98$.
As above, we show a Gaussian fit (dashed) to the data from the diffusive
dynamo (stars) driven by the ABC flow.}
\end{center}
\end{figure}
\begin{figure}
  \begin{center}
    \psfrag{z}[c]{$\log \zeta$}
    \psfrag{p}[c]{$\log P(\zeta)$}
    \includegraphics[width=0.4\textwidth]{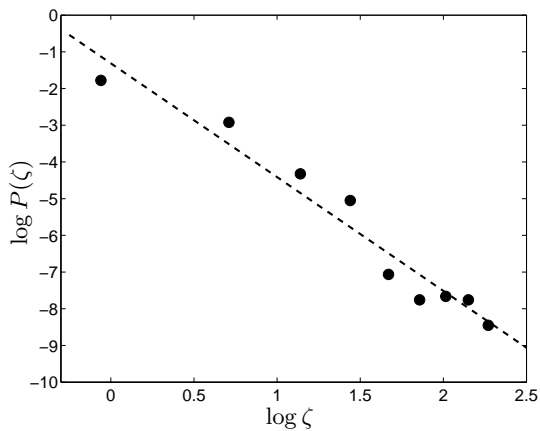}
    \caption{\label{pdf_sat} As in Fig.~\protect\ref{pdf_KS}, but for the
energy release rates from a nonlinear simulation (see
Fig.~\protect{\ref{saturated_brms}}). A power law fit is shown with dashed
line (with slope $-3.1$)}
\end{center}
\end{figure}

We show in Figs.~\ref{pdf_KS} and \ref{pdf_ABC} the probability distributions
of the magnetic energy release rate, normalised
to the total magnetic energy in the domain, $\zeta=(\gamma-\overline{\gamma})/\sigma_\gamma$, 
with $\gamma=\gamma_\mathrm{i}$ or $\gamma_\mathrm{r}$.
Here overbar denotes time averaging
(at times where an eigensolution has been established) and $\sigma_{\gamma}$ is the standard deviation of $\gamma$.
Since $\gamma=\Delta M/(M\tau)$, it can easily be seen that  
$\zeta=(\Delta M-\overline{\Delta M})/\sigma_{\Delta M}$.
We obtained the probability
distributions of $\zeta$ from both the induction equation
and the reconnection dynamo model, both driven by the KS flow, shown in
Fig.~\ref{pdf_KS}, and also with both based on the 111 ABC flow, shown in
Fig.~\ref{pdf_ABC}. The power-law index obtained for the KS flow is
$s\approx-3.3$, and that for the ABC flow is $s\approx-3.0$ . We stress that
this power-law behaviour is not related to the nature of the velocity field:
solutions of the induction equation with the same velocity fields, as we show
in Fig.~\ref{pdf_KS}, exhibit an approximately Gaussian probability
distribution.

Results shown in Figs.~\ref{pdf_KS} and \ref{pdf_ABC} have been obtained from
kinematic simulations, where the velocity field was not affected by the
Lorentz force. However, the corresponding nonlinear model introduced in
Section~\ref{NLM} retains this feature, with $s\approx -3.1$ for the KS flow in
the statistically steady state, as we show in Fig.~\ref{pdf_sat}.

It is not quite clear if the flux rope dynamo represents a physical
example of self-organised criticality, but the system seems to possess at
least some of the required properties. In particular, as we argue above, our model can be viewed as an extreme
case of magnetic hyperdiffusivity, which also arises in the self-organised
criticality models of Solar flares.

\section{Discussion and conclusions}
To summarise, we have confirmed that the dynamo action is sensitive to the
nature of magnetic dissipation and demonstrated that magnetic reconnections
(as opposed to magnetic diffusion) can significantly enhance the dynamo
action. We have explored the kinematic stage of the fluctuation dynamo in a
chaotic flow that models hydrodynamic turbulence and in the ABC flow, with the
only magnetic dissipation mechanism being the reconnection of magnetic lines
implemented in a direct manner. In our model, where magnetic dissipation is
suppressed at all scales exceeding a certain scale $d_0$, the growth rate of
magnetic field exceeds that of the magnetic diffusion-based fluctuation dynamo
with the same velocity field. Even when the velocity field of the
reconnection-based dynamo is reduced in magnitude as to achieve similar growth
rates of magnetic energy density, the rate of conversion of magnetic energy
into heat in the reconnection dynamo is an order of magnitude larger than in
the corresponding diffusion-based dynamo. Thus, reconnections more efficiently
convert the kinetic energy of the plasma flow into heat, in our case with the
mediation of the dynamo action. This result, here obtained for a kinematic
dynamo, can have serious implications for the heating of rarefied, hot plasmas
where magnetic reconnections dominate over magnetic diffusion (such as the
corona of the Sun and star, galaxies and accretion discs).

It is intriguing that reconnections play the same role \citep{Leadbeater:2001,Barenghi:2008} of converting kinetic energy into heat in superfluids and Bose-Einstein condensates, fluids near absolute zero at the opposite end of the temperature spectrum.

Our model can be viewed as a numerical implementation of the elusive limiting
regime of infinitely large magnetic Reynolds number, where magnetic
dissipation can be safely neglected at all large scales but plays a
crucial role at a certain very small scale (we are grateful to Alex
Schekochihin for suggesting this idea).

In contrast to the fluctuation dynamo based on magnetic diffusion, the
probability distribution function of the energy released in the flux rope
dynamo has a power law form not dissimilar to that observed for the Solar
flares. This is also true for the nonlinear states of the dynamo.

The reconnection-based dynamo model suggested here can be generalised to
include the modification of the velocity field by the Lorentz force. More
precisely, magnetic pressure is assumed to be balanced by the gas pressure, so
that only magnetic tension needs to be explicitly included into the
Navier--Stokes equation. Magnetic tension can readily be calculated
in our model where magnetic field is defined only at discrete positions of
closed magnetic loops. We suggest two approximations for the Navier--Stokes
equation, one designed to model Alfv\'en waves and the other suitable for the
studies of nonlinear dynamos. The former model can be useful in the studies of
nonlinear interaction of Alfv\'en waves and Alfv\'enic turbulence.

Unlike most -- if not all -- other simulations of the fluctuation dynamo, our
computations start with a spatially localised initial magnetic field. This has
allowed us to observe that the magnetised region spreads during the kinematic
dynamo stage but its size stops growing in the nonlinear stage. This can be
naturally interpreted as the suppression of the turbulent magnetic diffusion in
the saturated dynamo state. This is broadly equivalent to the reduction of
the effective magnetic Reynolds number down to its marginal value (with
respect to the dynamo action).

Our model of magnetic field evolution, based on tracing closed magnetic loops
can be fruitfully applied in other numerical approaches to
magnetohydrodynamics. One of well-known difficulties in the generalisation of
smoothed-particle hydrodynamics to include magnetic fields is the
implementation of the solenoidality of magnetic field. Quite notably, our
approach satisfies the magnetic solenoidality condition perfectly since
the modelled magnetic lines are closed at all times. A similar approach may be
fruitful in smoothed-particle magnetohydrodynamics codes.

\section*{Acknowledgements}
We thank Pat Diamond, Russell Kulsrud, Alex Sche\-ko\-chi\-hin, Andrew
Soward for useful discussions. This work was supported by the
STFC grant ST/F003080/1. AS is grateful to IUCAA for financial support and
hospitality.

\bibliographystyle{astron}
\bibliography{my}

\end{document}